%% file: AthermalUnfolding.tex
\documentclass[preprint,authoryear,3p,times]{elsarticle}
\input{Declarations.tex}

\journal{Journal of the Mechanics and Physics of Solids}
\begin{document} 


\title{Mechanics of collective unfolding }

\author[inria,upec]{M. Caruel}
\ead{matthieu.caruel@u-pec.fr}
\author[lms]{J.-M. Allain}
\author[lms]{L. Truskinovsky}
\address[inria]{Inria Saclay Ile-de-France, M3DISIM team, Palaiseau, France}
\address[upec]{
Université Paris Est, Laboratoire Modélisation et Simulation Multi Echelle, MSME-UMR-CNRS-8208, Bioméca - 61 Avenue du Général de Gaulle, 94010 Créteil, France
}
\address[lms]{
LMS,  CNRS-UMR  7649,
Ecole Polytechnique, Route de Saclay, 91128 Palaiseau,  France
}

\begin{abstract}
Mechanically induced unfolding of passive crosslinkers is a fundamental biological phenomenon   encountered across the scales from  individual macro-molecules  to  cytoskeletal actin networks. In this paper we study  a conceptual model of athermal load-induced  unfolding and use  a minimalistic setting allowing one to emphasize the role of long-range interactions while maintaining full analytical transparency. Our model can be viewed as  a description of a  parallel bundle of  $N$ bistable units confined between two shared  rigid backbones that are loaded through a series spring.  We show that the ground states in this  model correspond to synchronized, single phase  configurations where all individual units are either folded or unfolded. We then study the fine structure of  the wiggly  energy landscape  along the reaction coordinate linking the two coherent states and describing the  optimal mechanism of cooperative unfolding.  Quite remarkably, our study shows the fundamental difference in the size and structure of the folding-unfolding energy barriers in the hard (fixed displacements) and soft (fixed forces) loading devices which persists in the continuum limit. We argue that  both, the  synchronization and the non-equivalence of the mechanical responses in hard and soft devices, have their origin in the  dominance  of long-range interactions. 
We then apply  our minimal model to skeletal muscles where the  power-stroke in acto-myosin crossbridges can be interpreted as passive folding.  A quantitative analysis of the muscle model shows that the relative rigidity of myosin backbone provides  the long-range interaction mechanism allowing the system to effectively synchronize the power-stroke in individual crossbridges even in the presence of thermal fluctuations.
In view of the prototypical nature of the proposed model, our  general conclusions pertain to a variety of other biological systems where elastic interactions are mediated by effective backbones.
\end{abstract}

\maketitle

\section{Introduction}

In contrast to inert matter, distributed  biological systems are characterized by hierarchical network architectures with domineering long-range interactions.  Even in the absence of metabolic fuel  this leads to a highly nontrivial cooperative mechanical behavior in both statics and dynamics. Passive collective effects are usually revealed through synchronized conformational changes  interpreted here as generic folding-unfolding transitions. The experiment shows that such systems  exhibit coherent macroscopic hopping  between folded and unfolded configurations  resisting the destabilizing effect of finite temperatures \citep{ Dietz:2008gj, Thomas:2012ur, Erdmann:2013wr}. 

A \emph{minimal}  mechanical model showing the cooperative behavior is  a parallel bundle of bistable units linked by two shared backbones and its most natural biological prototype is a muscle  half-sarcomere undergoing the power-stroke  \citep{Caruel:2013jw}.  Another example is the  unbinding of focal adhesions with  individual adhesive elements  coupled through a common elastic background  \citep{Erdmann_2007}. Similar behavior has been also associated with mechanical denaturation  of RNA and DNA hairpins  where the long-range interactions are due  to the prevalence of stem-loop structures  \citep{Liphardt:2001fp,Bosaeus:2012kp}.  Other molecular systems with cooperative unfolding  include  protein $\beta$-hairpins \citep{Munoz:1998vf} and coiled coils \citep{Bornschlogl:2006we}.  The backbone dominated internal architecture  in all these systems leads to a \emph{mean-field} type mechanical feedback  which is also ubiquitous in bi-stable social systems \citep{Kometani_1975,Desai_1978}. 

In this paper we systematically study  the mechanics of the minimal model  in the setting which can be directly associated with  skeletal muscles.  We recall that the  mechanically induced  conformational change (power-stroke) in skeletal muscles takes place in  myosin heads (cross-bridges) that are bound in parallel to actin filaments \citep{Smith_2008,Linari_2010a,Guerin_2011,Erdmann:2012cr,Piazzesi:2014kr}.  The (thermo)mechanical behavior of this system was first analyzed  by Huxley and Simmons \citep{Huxley_1971} who interpreted  the  pre- and post-power-stroke conformations of the myosin heads as discrete chemical states (spin model). Similar  ideas have been independently advanced in the  studies of bistable adhesion clusters \citep{Bell_1978} and  in other  applications  ranging from Jahn-Teller effect and ripples in graphene sheets \citep{Bonilla:2012ib} to unzipping of biological macromolecules \citep{Gupta:2011cp,Prados:2012ks}.

Since in \citep{Huxley_1971}  the  behavior of the spin model was studied only  in a hard device (prescribed displacements),  the cooperative effects  were not found. To understand this surprising result we  study the  zero-temperature analog of the Huxley and Simmons  model.  We show  that  for this system the structures of the energy landscape in hard and soft  (prescribed forces)  loading conditions are rather different.  In particular, we explain  why in this model the collective behavior at finite temperature  can be expected in the soft but not in the hard device. To capture the coherent hopping in the hard device case, we regularize the spin model in two  ways.  First, following  \citep{Marcucci_2010} we replace the discrete chemical states (hard spins) by a continuous double-well potential with a finite energy barrier (snap-spring model). Second, to take into account the elastic interactions  between individual crosslinkers,  we introduce a series  spring mimicking the backbone elasticity \citep{Wakabayashi_1994,Huxley_1994} and bringing in mean-field interactions. 
We show  that in the snap-spring model the energy barriers separating the synchronized states are  still markedly higher in a soft device  than in a hard device which  provides an explanation for the observed retarded relaxation in isotonic experiments on skeletal muscles \citep{Reconditi_2004,Piazzesi_2002,Decostre_2005}.

An interesting peculiarity of  the snap-spring model is that the relaxed potential, representing the global minimum of the energy, is always convex in a soft device but is only convex-concave in a hard device. This means that the macroscopic stiffness,  which  is always positive in a soft device, can become negative in a hard device. The negative stiffness (metamaterial) response \citep{Nicolaou:2012cf}, which persists in the continuum limit, clearly contradicts the intuition developed in the studies of systems with short-range interactions. The non-convexity of the ground state energy in a hard device means that the system is non-additive and cannot be relaxed through the mixing of folded and unfolded units. 

To illustrate our general results, we consider in some detail the special case of  skeletal muscles where we can make quantitative estimates by using realistic parameters. Our analysis shows that the height of the microscopic energy barriers for the power-stroke  in individual cross-bridges  is of the order of the energy of thermal fluctuations. This implies that the cross-bridges can undergo conformational changes in a non-cooperative stochastic manner. However, the presence of  long-range interactions creates a bias in the individual folding-unfolding equilibria  in the form of a macroscopic barrier on top of which the microscopic barriers are superimposed.  We call this barrier macroscopic because its height  is proportional to the number of elements in the system. Due to the presence of the macroscopic barrier, the  folding transitions in individual cross-bridges become energetically preferable only after the top of this barrier has been reached which  means that individual cross-bridges have been synchronized.  These observations suggest that the elementary contractile unit of skeletal muscles has evolved  to control  the state of a large assembly of folding elements through a mean-field type mechanical feedback. Due to such passive synchronization, the power-stroke  takes place collectively which obviously amplifies the mechanical effect.

While we focus in this paper on the athermal behavior, our analysis reveals the origin of the anomalous thermodynamics and kinetics of skeletal muscles and similar systems observed at finite temperatures  \citep{Caruel:2013jw}. A detailed study of the temperature effects on the collective unfolding  will be presented elsewhere.
 
The paper is organized as follows. In Section \ref{sec:the_hs_model}, we study the equilibrium mechanical behavior of the spin model and show that already in this minimal setting the behavior of the system in  soft and hard devices is different. The snap-spring model is introduced in Section \ref{sec:the_SS_model} where we demonstrate that it  removes the degeneracies of the spin model and effectively interpolates between the soft and hard device behaviors. In the same section, we also study the fine  structure of the energy landscape separating the coherent states of the system and introduce a reaction coordinate to describe the successive individual folding-unfolding transitions constituting the collective unfolding. The adaptation of the snap-spring model for skeletal muscles is presented  in Section \ref{sec:application_to_the_case_of_skeletal_muscles}.  
Finally  in Section \ref{sec:conclusion} we present our conclusions.

\section{The hard spin (HS) model} 
\label{sec:the_hs_model}

\begin{figure}[htbp]
	\centering
	\includegraphics{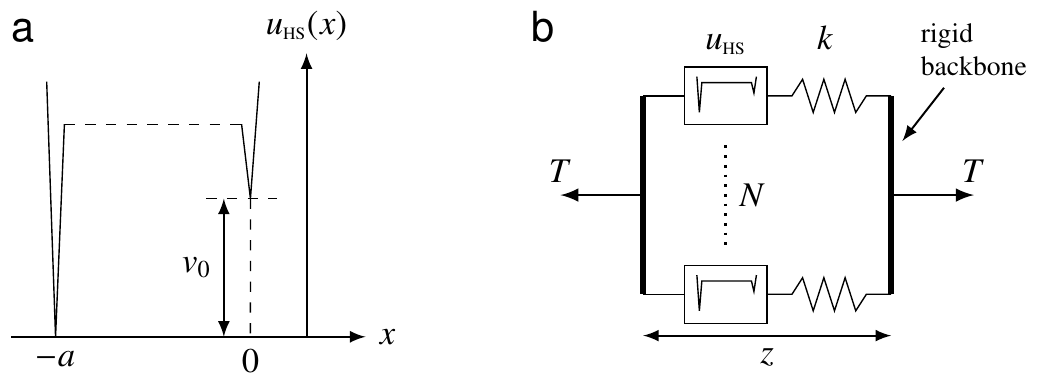}
\caption{Hard spin model of a parallel bundle of bistable crosslinkers.  (a) Energy landscape of an individual crosslinker; (b)  \( N \) crosslinkers loaded in a soft device.}
	\label{fig:hs_model}
\end{figure}

Consider the behavior of an elementary cluster of \( N \) bistable units connecting two rigid backbones.
In the spin model \citep{Huxley_1971}, each crosslinker is represented by a bistable potential connected to a series spring. The potential, representing two folding configurations, is assumed to have infinitely narrow energy wells representing two  chemical states,  see Fig.~\ref{fig:hs_model}. The potential describing individual spin units can be written in the form
\begin{equation}
\label{eq:HS_u}
 u_{\mbox{\tiny  HS}}\of{x} =
\begin{cases}
 v_{0}  & \text{if $ x= 0 $,}\\
0 & \text{if $ x=-a $.}
\end{cases}
\end{equation}
Here the spin variable $x$ takes two values, \( 0 \) and \( -a \),  describing the unfolded and the folded conformations, respectively. By $a$  we denoted the ``reference'' size of the conformational change interpreted as the  distance between the two  energy wells.  With the unfolded state we associate an energy level \( v_{0} \) while the folded configuration is considered as a zero energy state. The potential \eqref{eq:HS_u} is shown schematically in Fig.~\ref{fig:hs_model}(a).

\begin{figure*}[htbp]
	\centering
		\includegraphics[]{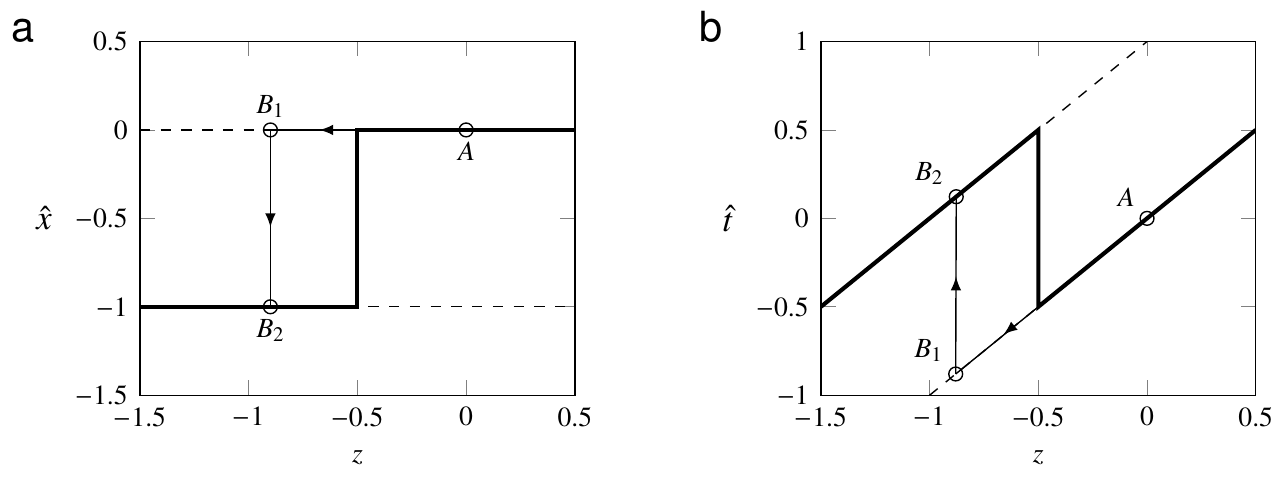}
	\caption{  Behavior of a  single crosslinker in the spin model. (a) Equilibrium states for various \( z \); (b) Corresponding tension levels. Dashed lines, metastable states; bold line, global minimum. Arrows show the  response to sudden shortening including a frozen elastic phase (\( A\to B_{1} \)) and a subsequent phase equilibration (\( B_{1}\to B_{2} \)). Here we used \( v_{0}=0 \).}
	\label{fig:HS_single_xb_equilibrium}
\end{figure*}

In addition to a spin unit with energy \eqref{eq:HS_u} each cross-bridge contains a linear shear spring with stiffness \( \k \); see Fig.~\ref{fig:hs_model}(b). The energy of the elastic spring is
\(
u_{\mbox{\tiny  E}}(x)=\k x^{2}/2
\)
and  the  energy of the whole crosslinker is
\begin{equation}
\label{eq:HS_u1}
u= u_{\mbox{\tiny  HS}}(x)+u_{\mbox{\tiny  E}}(z-x).
\end{equation}
Without loss of generality, we can assume that the reference length of the linear spring is already incorporated into the definition of the elongation \( z \). Notice that the   mechanical system with energy \eqref{eq:HS_u1}  has a  multi-stable response, see Fig.~\ref{fig:HS_single_xb_equilibrium}. 

To non-dimensionalize the resulting model, which one can  associate with the names of Huxley and Simmons  even though they never considered such parallel connection explicitly (for this representation, see \cite{Marcucci_2010}), we  choose \( a \) as a characteristic length,  associate the characteristic energy scale with   \( \k a^{2} \) and normalize forces by \( \k a \). The only remaining dimensionless parameters of the model are $N$ and \( v_0 \) and we can  write the dimensionless energy of the system (per crosslinker) in the form
\begin{equation}\label{eq:HS_v}
	v(\boldsymbol{x};z) = \frac{1}{N} \sum_{i=1}^{N} \left[(1+x_{i})\,v_0 + \frac{1}{2}(z-x_{i} )^{2}\right].
\end{equation}
Here, for convenience, we preserved the same notations for non-dimensional quantities.

In a hard device  each crosslinker is exposed to the same total elongation  \( z \) and thus the  individual units are independent.
In the soft device case, where the  control parameter is the total tension \( T \),  the energy per crosslinker is
\begin{equation}\label{eq:hs_w}
 w (\boldsymbol{x},z;t)= v(\boldsymbol{x},z) - tz= \frac{1}{N} \sum_{i=1}^{N} \left[(1+x_{i})\,v_0 + \frac{1}{2}(z-x_{i} )^{2}-tz\right],
\end{equation}
where \( t=T/N \) is the force per crosslinker. Now for each crosslinker both $x_{i}$ and $z$ are internal degrees of freedom and the individual units can no longer be considered as independent. Indeed, if we minimize out the global continuous variable $z$ by solving \( \left. \partial w/\partial z\right|_{t,\left\{x_{i}\right\}} = 0 \), we obtain
\begin{equation}\label{eq:hs_z_partial_eq}
	z= t + \frac{1}{N}\sum_{i=1}^{N}x_{i},
\end{equation}
which after inserting in Eq.~\ref{eq:hs_w} shows that the partially minimized energy depends quadratically on $\sum_{i=1}^{N} x_{i}$,
\begin{equation*}
	\tilde{w}(\boldsymbol{x};t) = \frac{1}{N}\sum_{i}\left[
		\frac{x_{i}^{2}}{2} - tx_{i} + (1+x_{i})v_{0} - \frac{t^2}{2} 
	\right]
	-\frac{1}{2}\left(\frac{1}{N}\sum_{i}x_{i}\right)^{2}.
\end{equation*}
One can see that the transition from hard to soft device introduces a mean-field interaction among the crosslinkers which, as we show later,  is ultimately responsible for the cooperative behavior.

\subsection{Mechanical equilibrium in a hard device} 
\label{sub:hs_hard_device}

To describe the equilibrium response of the HS model in a hard device, we need to compute the local minima of the mechanical energy \eqref{eq:HS_v} at fixed \( z \). 
Since each of the internal degrees of freedom \( x_{i} \), for \( 1 \leq i \leq N \) can take only two discrete values, \( x_{i}=\hat{x}_{0} =0 \) and \( x_i =\hat{x}_{1} = -1 \), a given equilibrium state is characterized by the distribution of the \( N \) crosslinkers between the two spin configurations.
Due to the permutational invariance of the problem, each equilibrium state is fully characterized by a discrete  order parameter representing the fraction of crosslinkers in the folded state,
\begin{equation}\label{eq:hs_p}
p=\frac{1}{N}\sum_{i=1}^{N}\alpha_{i},
\end{equation}
where \( \alpha_{i}=1 \) if \( x_{i}=-1 \) and \( \alpha_{i}=0 \) if \( x_{i}=0 \).  
\begin{figure*}[!ht]
	\centering
	\includegraphics{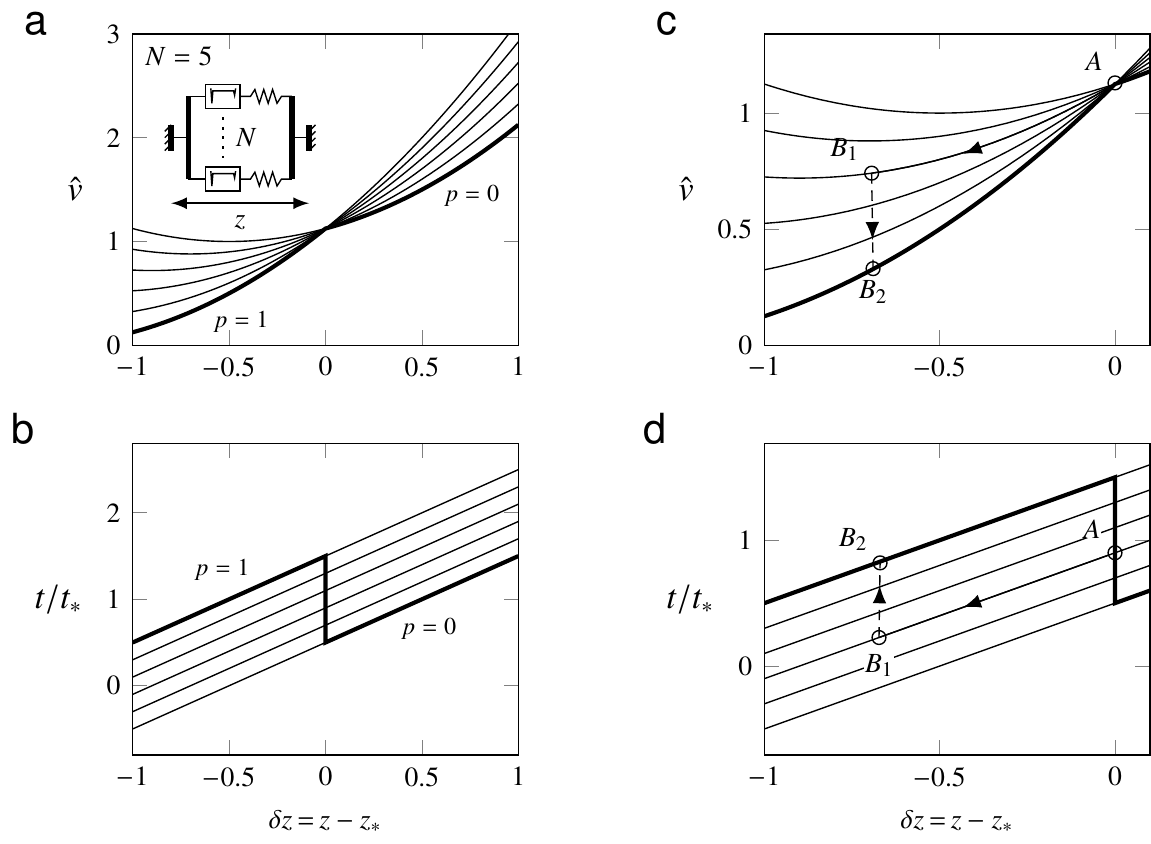}
	\caption{Mechanical response of the HS model in a hard device with \( N=5 \). (a) Energy levels of the metastable states (\( p=0,\frac{1}{5},\dots,1 \)) for different applied elongations. (b) corresponding tension-elongation relations. (c) and (d) are details of (a) and (b) with an illustrated response path to a fast loading experiment including a frozen elastic phase (\( A\to B_{1} \)) followed by a subsequent phase equilibration (\( B_{1}\to B_{2} \)). Thick lines, global minimum corresponding to \( p=0 \) (resp. \( p=1 \)) for \( z>z_* \) (resp. \( z<z_* \)) . \( t_* = v_0 \) and \( z_*=v_0-1/2 \). Parameters are \( v_{0}=1 \) and \( N=5 \).}
	\label{fig:HS_zero_temperature_hard_device}
\end{figure*}

In \ref{app:hard_device} we show that all equilibrium configurations of this type  correspond to   local minima of the energy \eqref{eq:HS_v}. At a given value of the order parameter \( p \), the energies of all such  metastable states are equal to
\begin{equation}
	\hat{v}(p;z) =  p\frac{1}{2}\left( z+1 \right)^{2} + (1-p)\bigl( \frac{1}{2}z^{2} + v_{0} \bigr).\label{eq:hs_hd_energy_local_minima}
\end{equation}
This is a  linear combination of the energies of two limiting configurations, one fully folded with $p = 1$ and  the  energy \( \frac{1}{2}(z+1)^{2} \) and the other one fully unfolded with \( p=0 \) and  the  energy \( \frac{1}{2}z^{2} + v_{0} \).
The absence of  a \emph{mixing} energy is a manifestation of the fact that the two coexisting populations of crosslinkers do not interact.

The tension-elongation relations along metastable branches parameterized by $p$ can be written as
\begin{equation}
\hat{t}(p;z)= \frac{\partial}{\partial z}\hat{v} (p;z)= z+p,\label{eq:hs_hd_tension_local_minima}
\end{equation}
so for given \( p \)  we obtain equidistant parallel lines, see Fig.~\ref{fig:HS_zero_temperature_hard_device}. A peculiar feature of the HS model is that the domain of hysteretic behavior extends indefinitely because the spin system does not have any stress thresholds.

To find the global minimum of the energy we need to perform at each value of $z$ an additional minimization over the discrete variable $p$.   If we compute the derivative
\[
 \frac{\partial}{\partial p}\hat{v}(p;z) = z+\frac{1}{2}-v_{0}.
\]
which does not depend on \( p \), we obtain that for \( z>z_{*}\), where \( z_{*} = v_{0} - 1/2 \), the global minimizer is the folded state  with \( p=1 \) and for \( z<z_{*}  \)  it is  the unfolded state with \( p=0 \), see   Fig.~\ref{fig:HS_zero_temperature_hard_device}. The global minimum energy profile exhibits a  kink near the crossing (folding) point. Similar kinks associated with unfolding of hairpins and other folding patterns have been observed in the energy profiles reconstructed from single molecule force spectroscopy  measurements of proteins and nucleic acids \citep{Gupta:2011cp}.

Observe that  the ground state energy  in this model is \emph{nonconvex} independently of the number of units. This means that the energy is not  convexified through the formation of mixtures as in systems with \inserted{short-range} interactions (see, for instance, \cite{Puglisi_2000}). The reason is that this mean-field system is strongly non-additive and all mixed states are energetically unfavorable due to high cost of mixing. Somewhat similar situation takes place  in  theory of elastic phase transitions where the relaxed energy is also nonconvex in the general case (it is only quasi-convex) which is again the consequence of long-range elastic interactions exemplified by the gradient constraint  (see, for instance, \cite{Ball:2002dr}).

A consequence of the energy nonconvexity is the non-monotonicity of the force-elongation relation  shown  in  Fig.~\ref{fig:HS_zero_temperature_hard_device}(b,d). More specifically, the system exhibits a \emph{negative}  stiffness at the point where all crosslinkers collectively flip from the folded to the unfolded state.  Similar   mechanical behavior has been recently artificially engineered in \emph{metamaterials} by drawing on the Braess paradox for decentralized  globally connected networks \citep{Cohen:1991gm,Nicolaou:2012cf}. As in our case,  the mean-field type coupling in  metamaterials is achieved via parallel connections with multiple shared links. 
Biological examples of  systems with negative stiffness are provided by unzipping  RNA and DNA hairpins \citep{Woodside:2008uz,Bosaeus:2012kp}.

\subsection{Mechanical equilibrium in a soft device} 
\label{sub:hs_soft_device}
Consider now the HS model loaded in a soft device when the equilibrium states
correspond to local minima of the mechanical energy  \eqref{eq:hs_w}. An equilibrium state is again fully characterized by the fraction of units in the folded state  \( p \), defined by Eq.~\ref{eq:hs_p}.
 We can then write the (marginal) energy of the partially equilibrated system as
\begin{equation*}
	\tilde{w}(p,z;t) = \hat{v}(p,z) - tz,
\end{equation*}
where \( \hat{v} \) is the energy of the metastable branch parametrized by \( p \) in the hard device case, see Eq.~\ref{eq:hs_hd_energy_local_minima}.

Then we eliminate  \( z \) by using Eq.~\ref{eq:hs_z_partial_eq}.
Each value of the order parameter \( p \) defines a branch of local minimizers of the energy \eqref{eq:hs_w} parameterized by $t$, see \ref{app:soft_device}. At a given value of \( p \), the energy of a metastable state reads
\begin{equation}
\hat{w}(p;t) = -\frac{1}{2}t^{2} + pt + \frac{1}{2}p(1-p) + (1-p)v_{0}. \label{eq:hs_sd_energy_local_minima}
\end{equation}

\begin{figure*}[!ht]
	\centering
	\includegraphics{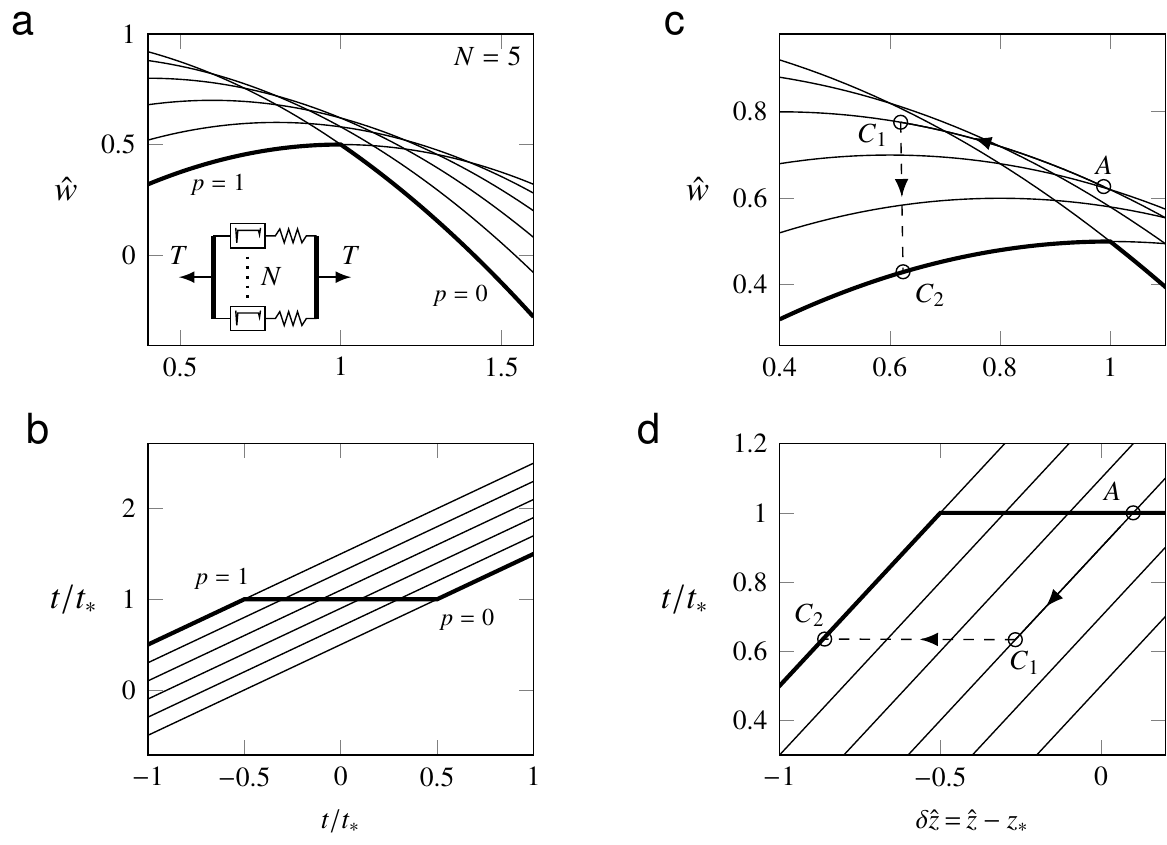}
	\caption{ Mechanical response of the HS model in a soft device. (a) Energy levels of the metastable states (\( p=0,\frac{1}{5},\dots,1 \)) for different applied forces. (b) Corresponding tension-elongation relations. In   (c) and (d) we zoom into    domain near \( t = t_* = v_0 \) and show schematically the response of the system to a sudden application of a load increment with an elastic phase (\( A\to C_{1} \)) followed by a folding phase \( C_{1}\to C_{2} \). Thick lines, global minimum corresponding to \( p=0 \) (\( p=1 \))  for \( t>t_* \)  (\( t<t_* \)).  Parameters are, \( v_{0}=1 \) and \( N=5 \).	}
	\label{fig:HS_zero_temperature_soft_device}
\end{figure*}
In contrast to the case of a hard device, here there is a nontrivial coupling term \( p(1-p)\) describing the energy of mixing. The presence of this term is  a signature of a mean-field  interaction among individual crosslinkers. Indeed, if one element changes configuration, its contribution to the common tension changes accordingly and the other elements must adjust to maintain the force balance. The tension-elongation relation associated with a set of metastable states sharing the same value of the parameter \( p \) can be written in the form
\begin{equation*}
\hat{z}(p;t) = -\frac{\partial}{\partial t}\hat{w} (p;t)=t - p.
\end{equation*}
At a given value of $p$ this relation is identical with its counterpart in the case of a hard device, see Eq.~\ref{eq:hs_hd_tension_local_minima}.
The globally stable states  can be found by minimizing \eqref{eq:hs_sd_energy_local_minima} over \( p \). Since \( \partial^{2}\hat{w}/\partial p ^{2} = -1 \), this function is  concave in \( p \). Therefore, the global minimum is again attained either at \( p=1  \) or  \( p= 0 \).
The energies of the two coherent configurations with \( p=1 \) and \( p=0 \) coincide when \(t = t_{*} = v_{0}\).

Our Fig.~\ref{fig:HS_zero_temperature_soft_device} illustrates  the structure of the energy-tension and the tension-elongation relations corresponding to different values of \( p \)  for the system with \( N=5 \). Notice that in contrast to the case of a hard device, the force-elongation relation characterizing the global minimum in  a soft device exhibits a \emph{plateau} indicating a discontinuity in elongation as the crosslinkers switch collectively at \( t = t_* \) from unfolded to folded conformation. 

Notice that the plateau replaces the  region of  negative stiffness detected in the hard device case and the force-elongation relation becomes monotone. This shows that even in the continuum limit  the stable ``material'' responses of our system in  hard and soft devices remain different. Such behavior would be completely unexpected from the perspective of classical statistical mechanics,  however, it is characteristic of systems with domineering long-range interactions \citep{Dauxois:2003umb}. 
 
\subsection{Energy landscape} 
\label{sub:hs_energy_barriers}

As we have seen in the previous sections, the globally stable state of the HS system correspond to one of the two coherent configurations characterized by \( p=1 \) and by \( p=0 \). We can now pose the question about the size of the energy barrier separating these two configurations.
To access the energy barriers  and to find the transition states (saddle points of the energy) we study the energy dependence on \( p \).
For general values of the loading parameters this dependence was found to be \emph{linear} in a hard device, indicating that there is no conventional barrier, and \emph{concave} in a soft device which means that there is a potential energy barrier. 
This observation shows that a switch from unfolded to folded configuration  in a soft device carries an  energetic cost while in a hard device the transition is cost-free.
\begin{figure*}[htbp]
	\centering
		\includegraphics{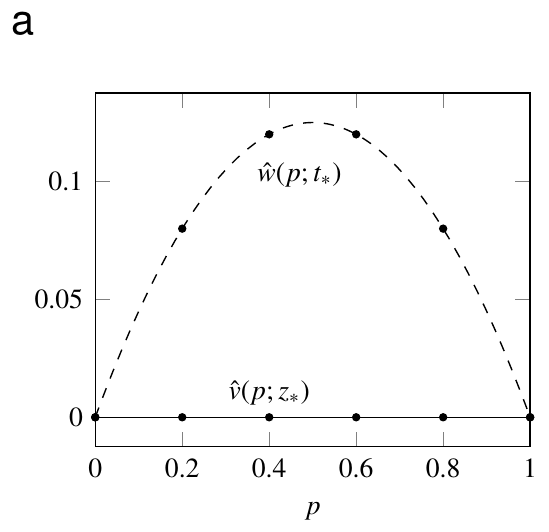}
	\hspace{1cm}	\includegraphics[]{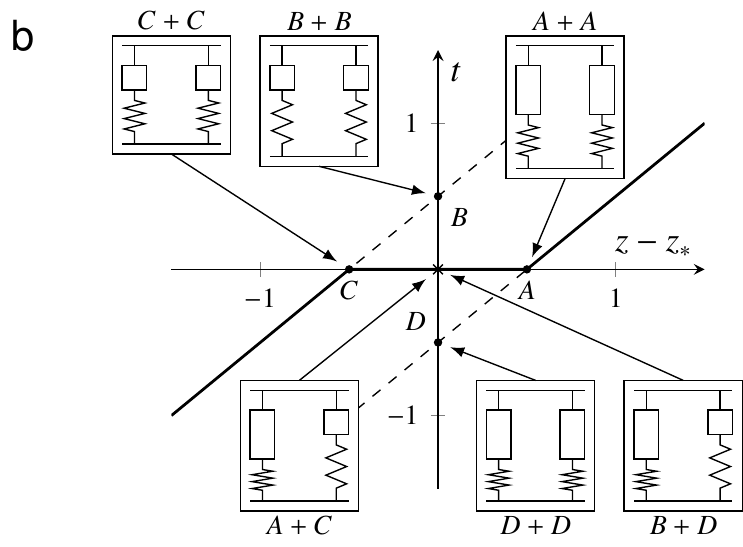}
	\caption{(a) Energy landscape  at the global minimum transition for the HS model. Solid lines, hard device at \( z=z_{*} \); Dashed lines, soft device at \( t=t_{*} \). Dots represent the energy of the different configurations for a system with \( N=5 \); lines correspond to the limit \(N \to \infty\).  Parameters are \( v_0=1 \) and \( N=5 \). (b) Representation of the behavior of a system with two crosslinkers with \( v_0=0 \) imposing \( t_*=0 \) and \( z_{*}=-1/2 \), the transition of the global minimum tension-elongation curve (thick line) occurring in a stress-free configuration in a soft device. Dashed lines, metastable states \( p=0 \) and \( p=1 \). The intermediate stress-free configuration is obtained either by mixing the two geometrically compatible states \( B \) and \( D \) in a hard device which results in a \( B+D \) structure without additional internal stress or by mixing the two geometrically incompatible states \( A \) and \( C \) in a soft device which results in a \( A+C \) structure with internal residual stress.}
	\label{fig:HS_zero_temperature_transition}
\end{figure*}

The (collective) transition takes place  at \( z = z_{*} \) in a hard device and \( t=t_{*} \) in a soft device. The difference between the corresponding energy landscapes at the threshold values of parameters  is illustrated in Fig.~\ref{fig:HS_zero_temperature_transition} (a). For the ease of comparison  the energy minima are shifted to zero in both loading conditions.
 Each black dot represents the energy level of a particular metastable state  and the dotted line represents the set of such metastable states in the continuum limit \( N\to\infty \) when \( p \) becomes a continuous variable. 
The barriers separating individual metastable states are not defined in the spin model. A simple way to  recover  microscopic barriers by switching from hard to soft spins  is discussed in Sec.~\ref{sec:the_SS_model}; for another regularization approach see \citep{Benichou:2013eo}.
 
To understand the origin of this peculiar behavior of the energy landscape it is instructive to  consider the minimal HS system with \( N=2 \); see Fig.~\ref{fig:HS_zero_temperature_transition} (b). Here  for simplicity we assumed that \( v_0=0 \) implying  \( t_*=0 \) and \(  z_{*}=-1/2 \). The two ``pure'' configurations are labeled as \( A \) (\( p=0 \)) and \( C \) (\( p=1 \))  at \( t=t_*=0 \) and as \( D \) (\( p=0 \)) and \( B \) (\( p=1 \))  at \( z=z_*=-1/2 \). In a hard device, where the two elements do not interact, the transition from state \( D \) to state \( B \) at a given $z=z_{*}$ goes through the configuration \( B+D \) which has the same energy as configurations \( D \) and \( B \): the crosslinkers in folded and unfolded  states are geometrically perfectly compatible and their mixing requires no additional energy.
 Instead, in a soft device, where individual elements interact, a transition from state \( A \) to state \( C \) taking place at a given \( t=0 \) requires passing through the transition state \( A+C \) which has a nonzero residual stress. Individual crosslinkers in this mixture state have different  values of \( z \) and therefore the energy of the stressed ``mixed'' configuration \( A+C \) is larger than the energies of the ``pure'' unstressed states  \( A \) and \( C \).
 We conclude that the macroscopic barrier in a soft device is higher than in a hard device  because in a soft device a transition is a genuinely cooperative effect requiring essential interaction of individual elements while  in a hard device the conformational change in different units takes place independently.
 
  \begin{figure}[!ht]
 	 \centering
 	 \includegraphics{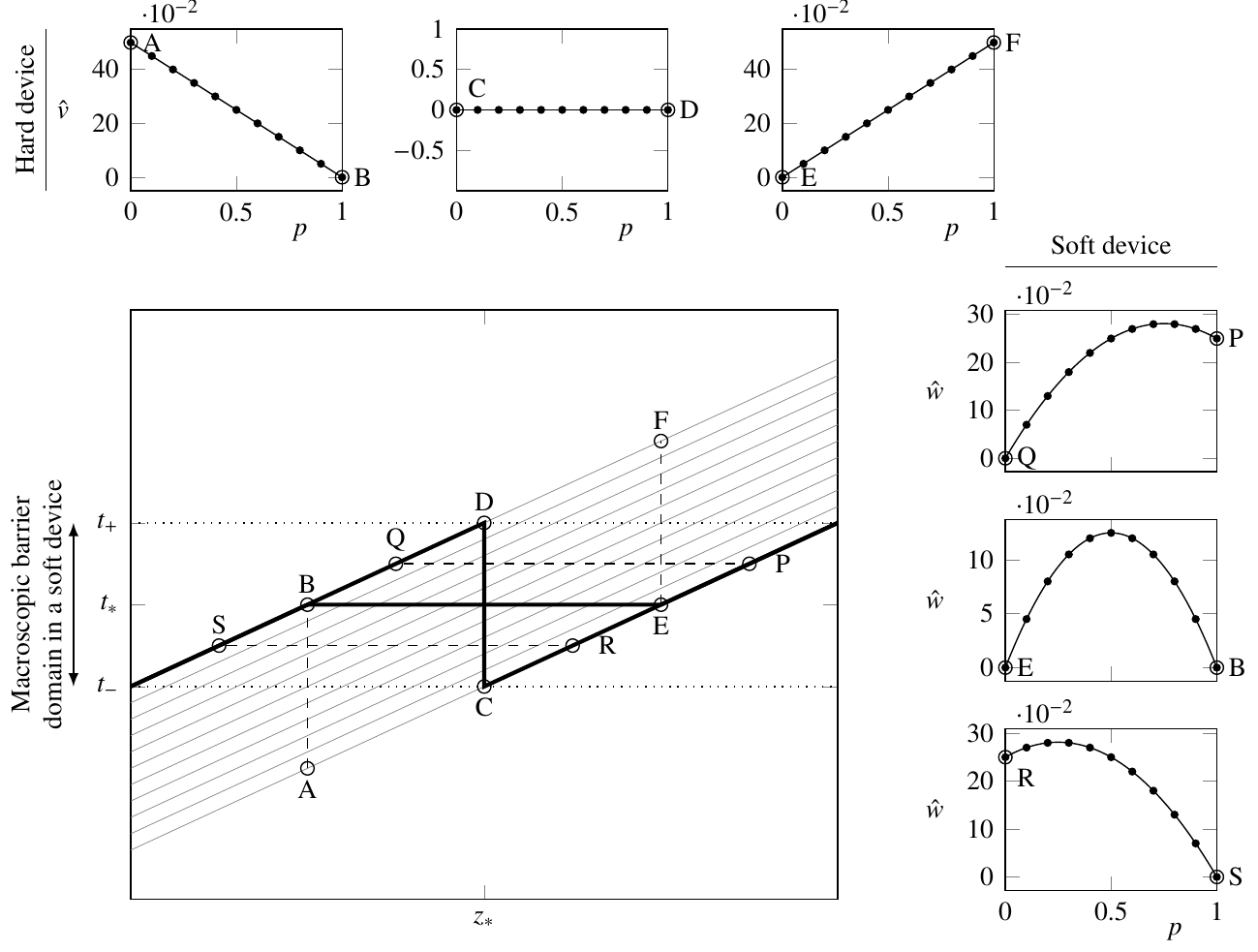}
  	\caption{\label{fig:hs_summary}Summary of the behavior of the HS model. The main frame shows the tension-elongation relations corresponding to \( p=0,0.1,\dots,1 \) for a system with \( N=10 \) (gray lines) and the tension-elongation relation in the global minimum (thick lines). The horizontal dotted lines show the limit of the domain where a macroscopic energy barrier is present in a soft device. The satellite frames (above, hard device; right, soft device) show the energy barriers corresponding to various transitions (A\( \to \)B, C\( \to \)D,...) shown by dashed lines in the main frame.}
  \end{figure}
  
 For general values of the loading parameters, we consider the \( N\to\infty \) limit where the energy landscape becomes smooth and  the barrier is located at the saddle point   \( p=p_{*} \), where \( p_{*}=t-v_0+1/2 \) is
  the solution of \( \partial\hat{w}/\partial p = 0 \).
  Since \( 0\leq p_{*}\leq 1 \), we obtain that the macroscopic barrier and the ensuing cooperative effects exist for \( t_-\leq t \leq t_+ \) with \( t_{-}=t_{*}-1/2 \) and \( t_{+}=t_{*}+1/2 \), in the soft device setting. Notice that the boundaries of this interval correspond to the threshold \( \hat{t}(0;z_*) \) and \( \hat{t}(1;z_*) \) delimiting the region with negative stiffness in the hard device case, see Fig.~\ref{fig:HS_zero_temperature_hard_device}.

We summarize the results obtained within the HS model in Fig.~\ref{fig:hs_summary}. In the main frame, we show the tension-elongation relations corresponding to \( p=0,0.1,\dots,1 \), for a system with \( N=10 \) (gray lines). Thick lines show the tension-elongation relations in the global minimum characterized by a transition located at \( z_{*} \) in a hard device and \( t_{*} \) in a soft device.
The horizontal dotted lines show the limits of the domain where a macroscopic barrier is present in a soft device.
 On the satellite frames (top - for a hard device and  right - for a soft device) we show the energy barriers between the homogenous configurations in the case \( N\to\infty \) (solid lines) and for discrete values of \( p \) (dots) at different values of the loading. 
 
To conclude, we have shown that the HS model exhibits different mechanical responses in a hard and a soft devices even though in both cases the global minimum of the energy is achieved for homogenous configurations
This behavior originates from the presence of long-range interactions between the crosslinkers. These interactions introduce an elastic feedback in the soft device case which creates a macroscopic energy barrier for the cooperative folding-unfolding process.

\section{The snap-spring model} 
\label{sec:the_SS_model}

The minimal description of unfolding transition provided by the spin model has its limitations. For instance,  the description of the barriers in the HS model is incomplete because it does not take into account the  microscopic energy barriers between the states with different values of the order parameter \( p \).  Another problem is that in the hard device setting the spin model is degenerate because configurations with different values of the order parameter are  equivalent.

\begin{figure}[htbp]
 \centering
\includegraphics{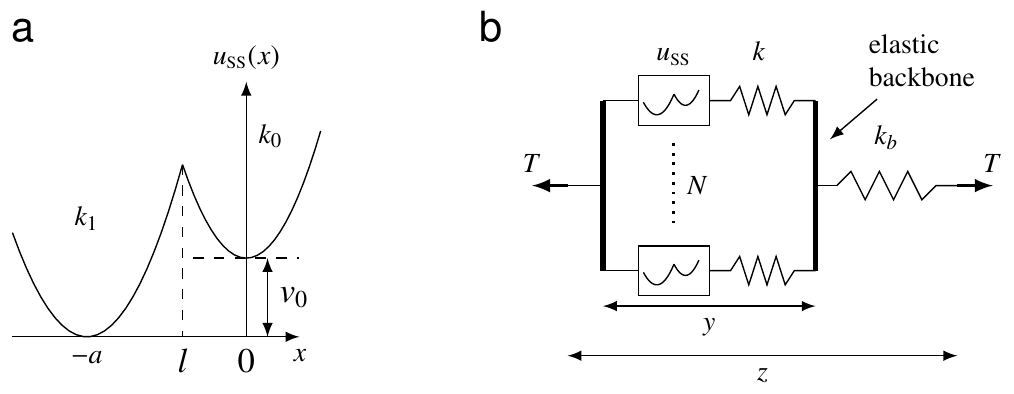}
 \caption{Snap-spring model of a cluster. (a) Dimensional energy landscape of a bistable crosslinker. (b) Structure of a parallel bundle  containing \( N \) crosslinkers in a soft device.}
\label{fig:cat_model}
\end{figure}

To deal with these problems we regularize the spin model by introducing two additional physical mechanisms. 
First, following  \citep{Marcucci_2010} we replace hard spins by snap-springs also known as soft spins, so that \( x \) becomes a continuous variable. For simplicity we assume that the corresponding double-well potential can be represented as a minimum of two parabolas, see Fig.~\ref{fig:cat_model}(a).
By using non-dimensional variables we can then write
\begin{equation*} \label{eq:u_SS}
 u_{\mbox{\tiny SS}}(x) =
\begin{cases}
 \frac{1}{2}k_0(x)^{2} + v_0   & \text{if $ x> l $}\\
  \frac{1}{2}k_1(x + 1)^{2}  & \text{if $ x\leq  l $}
\end{cases}
\end{equation*}
Here \( l \) is the dimensionless position of the energy barrier,  \( v_{0} \) is the dimensionless energy bias of the unfolded state and \( \k_{1} \) and \( \k_0 \), are dimensionless elastic moduli of the folded and unfolded states, respectively.  Interestingly,  a comparison with the reconstructed potentials for unfolding biological macro-molecules shows that this approximation may be in fact very good \citep{Gupta:2011cp}.
A spinodal region can be obviously added to the potential of the bi-stable unit, however, in this case we lose transparency without gaining essential effects.

\begin{figure*}[htbp]
	\centering
		\includegraphics[]{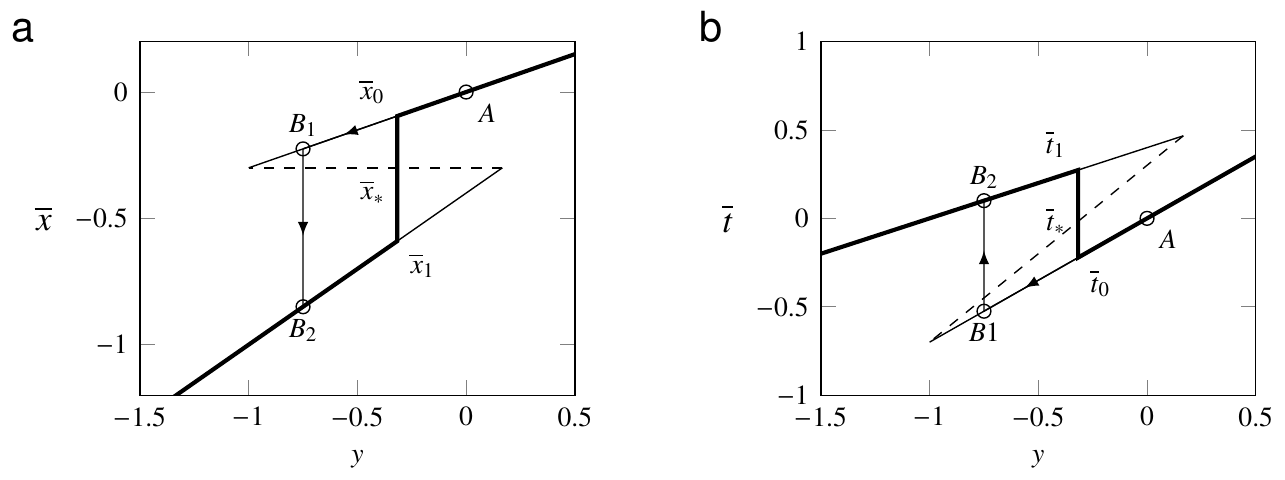}
	\caption{Behavior of a single crosslinker in the snap-spring model. (a) Equilibrium positions for various \( y \); (b) Corresponding tension levels. Solid lines, metastable states; dashed lines, unstable state; bold line, global minimum. Arrows indicate the response to a sudden shortening with an elastic unloading in the unfolded state (\( A\to B_{1} \)) followed by the conformational change to the folded state (\( B_{1}\to B_{2} \)). Parameters are, \( \lambda_{1} = 0.4 \), \( \lambda_0=0.7 \), \( l = -0.3 \)}
	\label{fig:SS_single_xb_equilibrium}
\end{figure*}

The important observation is that while  in the snap-spring model the bottoms of the energy wells remain the same as in the spin model,  the barrier separating the two conformational states is now well defined, see Fig.~\ref{fig:cat_model}. The mechanical response of a single  crosslinker with  an attached  series spring is governed by the dimensionless energy
\[
u=u_{\mbox{\tiny SS}}(x)  + \frac{1}{2}(y-x)^{2},
\]
where \( y \) is the total elongation. The solutions  of  the equilibrium equation  \( u'_{\mbox{\tiny SS}}(x) = y-x \) are shown in  Fig~ \ref{fig:SS_single_xb_equilibrium}. Notice the multi-valuedness of the relation linking  the variables \( x \) and \( y \) and the appearance of the   spinodal   branch \( \bar{x}_{*} \)   connecting the two stable branches \( \bar{x}_{1}(y) \) and \( \bar{x}_{0}(y) \).
 
The mechanical  independence  of the crosslinkers in a hard device disappears if we  take into account the  finite stiffness of backbone which, in the case of skeletal muscle, corresponds to the combined stiffness of actin and myosin filaments \citep{Wakabayashi_1994,Huxley_1994,Ford:1981wd,Mijailovich_1996,DeGennes_2001}. Following \citep{Julicher_1995}, we use a lump description of backbone elasticity by introducing an additional elastic spring  with stiffness \(\lambda_b=\k_{b}/(N\k)\) and the energy
\(
u_{b}(x) =  N\lambda_b x^2/2.
\)
If we attach this spring in series to our parallel bundle of crosslinkers, see Fig.~\ref{fig:cat_model}(b),  we can write the total energy of the system per crosslinker  in the form
\begin{equation}\label{eq:SS_v}
 v(\boldsymbol{x},y;z)=\frac{1}{N} \sum_{i=1}^{N} \left[u_{\mbox{\tiny SS}}(x_{i})+ \frac{1}{2}(y-x_{i})^{2} +\frac{\lambda_{b}}{2}(z-y)^{2}\right].
\end{equation}
In the hard device case \( z \) is the control parameter, \(x_i\) are the continuous microscopic internal variables generalizing the spin variables in the HS model and $y$ is a new continuous mesoscopic internal variable. Notice that now even in a hard device the individual crosslinkers are not independent; the implicit mean-field interaction  becomes obvious if the variable \( y \) is adiabatically eliminated (minimized out) by solving 
\(
 \left.\frac{\partial v}{\partial y}\right|_{z,{\boldsymbol{x}}}=0 
 \). 
 We obtain
 \begin{equation}\label{eq:SS_equilib_y}
 	\bar{y}({\boldsymbol{x}};z) = \frac{1}{1+\lambda_{b}}\left(\lambda_{b} z + \frac{1}{N}\sum_{i=1}^{N}x_{i}\right),
 \end{equation}
which after inserting into Eq.~\ref{eq:SS_v} gives the partially equilibrated energy
\begin{equation*}
	\tilde{v}(\boldsymbol{x};z) = 
	\frac{1}{N}\sum_{i}
	\left[
		u_{\mbox{\tiny SS}}(x_{i}) 
		+ \frac{x_{i}^{2}}{2}
		- \frac{\lambda_{b}z}{\left(1+\lambda_{b}\right)}x_{i}
		+\frac{\lambda_{b}z^{2}}{2\left(1+\lambda_{b}\right)}
	\right]
	-\frac{1}{2(1+\lambda_{b})}\left(\frac{1}{N}\sum_ix_{i}\right)^{2}.
\end{equation*}
Observe  that the quadratic term in \( \sum x_{i} \) vanishes when the elasticity of the backbone becomes infinite (\( \lambda_{b}\to\infty \))  showing that in this limit  the long-range interactions disappear.

The resulting snap-spring model can be viewed as a regularization of the HS model. To recover the HS model in a hard device case from Eq.~\ref{eq:SS_v}, we need to perform the  double limit: \(k_{1,0}\to\infty\) and \(\lambda_{b}\to\infty\). The first of these limits ensures that $x$ becomes a spin variable while the second  guarantees that \(y=z\).  To obtain the HS model in a soft device we need to consider the triple asymptotics: \(k_{1,0}\to\infty\), \( \lambda_{b}\rightarrow 0\) and \(z \rightarrow \infty\) where the last two limits  must be linked in the sense that  \( \lambda_{b}z \rightarrow t\) which ensures that  the  force per crosslinker \( t \)  remains finite.

In a soft device,  the total energy  per crosslinker  in the snap-spring model  can be written as
\begin{equation*}
 w(\boldsymbol{x},y,z;t) = v(\boldsymbol{x},y,z) -tz = \frac{1}{N} \sum_{i=1}^{N} \left[u_{\mbox{\tiny SS}}(x_{i})+ \frac{1}{2}(y-x_{i})^{2}+\frac{\lambda_{b}}{2}(z-y)^{2}- tz\right]   ,
\end{equation*}
where \(t=T/N \) is again the applied force per crosslinker.

\subsection{Mechanical equilibrium in a hard device} 
\label{sub:SS_hard_device}

To find equilibria in a hard device  we need to solve the following system of equations
\begin{subequations}\label{eq:SS_system}
	\begin{empheq}[left=\empheqlbrace]{align}\label{eq:SS_system_xi}
	\left.\frac{\partial v}{\partial x_{i}}\right|_{z,y,\left\{x_{j\neq i}\right\}} &= 0\quad
	\text{for all } 1\leq i \leq N\\
	\left.\frac{\partial v}{\partial y}\right|_{z,\left\{x_{i}\right\}} &= 0\label{eq:SS_system_y}
	\end{empheq}
\end{subequations}
Equation\ref{eq:SS_system_xi} have up to 3 solutions that can be parameterized by \( y \),
	\begin{equation}
	\begin{cases}
	\bar{x}_1\of{y} = \left(1-\lambda_1\right)y-\lambda_1, & \textrm{if\ }x_{i} < l\\
	\bar{x}_0\of{y} =\left(1-\lambda_0\right) y, & \textrm{if\ } x_{i} >  l\\
	\bar{x}_{*} = l&
	\end{cases}
	\label{eq:cat_ec1_ec0_def}
	\end{equation}%
where we  redefined the dimensionless parameters
\[
	 \lambda_{0} = \frac{\k_{0}}{1+\k_{0}},\quad
	\lambda_{1} = \frac{\k_{1}}{1+\k_{1}}.
	\]
	The solution of Eq. \ref{eq:SS_system_y} given by Eq.~\ref{eq:SS_equilib_y} allows to express the  mesoscopic variable $y$ through the microscopic variables $x_i$.
Because of the permutational invariance, the equilibrium solution of Eq.~\ref{eq:SS_system} is  fully characterized by the  fraction \( p \) of crosslinkers in the folded conformation \( \bar{x}_1\of{y} \)  and the fraction \( r \) of crosslinkers in the unfolded conformation \( \bar{x}_0\of{y} \).  The fraction of crosslinkers in the ``spinodal point''   \(  \bar{x}_{*} \) is then  $q=1-p-r$.

Using  \eqref{eq:SS_equilib_y}, we can eliminate the variable $y$ and obtain an explicit representation of the microconfiguration in terms of \( (p,q,r) \):
\begin{align}
	\hat{y}\of{p,q,r;z} &= \frac{\lambda_{b}}{\lambda_{b}+\Lambda(p,q,r)}\bigg(z-\frac{p\lambda_1-ql}{\lambda_{b}}\bigg)\label{eq:SS_hat_y_of_p_hd},\\
	\hat{x}_1\of{p,q,r;z} &= \frac{1-\lambda_1}{\lambda_{b}+\Lambda(p,q,r)}\left( \lambda_{b}z - p\lambda_1+ql\right)-\lambda_1\label{eq:SS_hat_x1_of_p_hd}, \\
	\hat{x}_0\of{p,q,r;z} &= \frac{1-\lambda_0}{\lambda_{b}+\Lambda(p,q,r)}\left( \lambda_{b}z - p\lambda_1+ql\right). \label{eq:SS_hat_x0_of_p_hd}
\end{align}
Here
\(
\Lambda(p,q,r) = p\lambda_{1} + q + r\lambda_{0}
\)
represents the equivalent stiffness of the parallel bundle of crosslinkers in a mixed configuration parameterized by \( (p,q,r) \).
\begin{figure*}[!ht]
	\centering
	\includegraphics{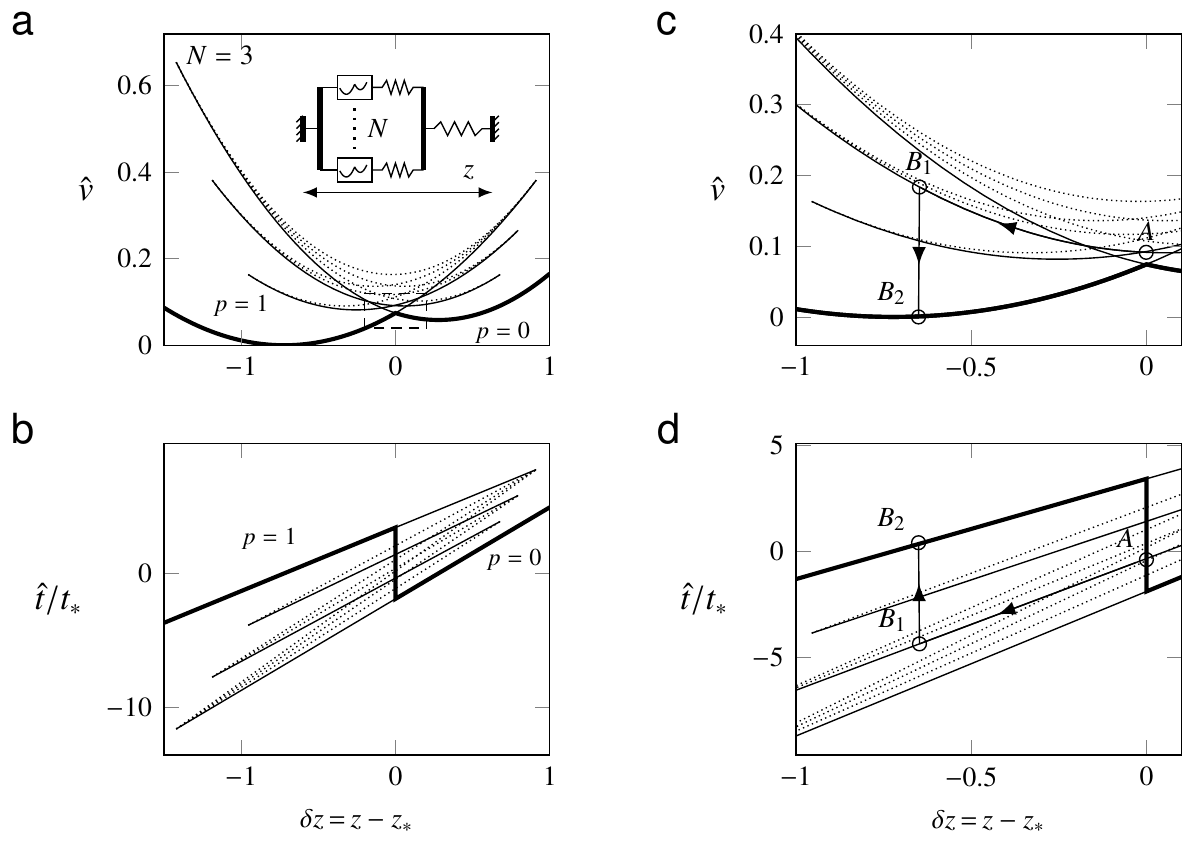}
	\caption{Mechanical response of the snap-spring model in a hard device. (a) Energy levels of all the \( (p,q,r) \) configurations for the case \( N=3 \) at different applied elongations. (b) corresponding tension-elongation relations. Solid lines, metastable states with \( p=0,1/3,2/3,1 \) and \( r=1-p \), \( q=0 \); dotted lines, unstable states with \( q\neq 0 \); thick lines, global minimum corresponding to \( p=0,\,r=1 \), for \( z>z_* \), and to \( p=1,\,r=0 \), for  / \( z<z_* \).
	(c,d) blow-up of (a,b) illustrating the response of the system to abrupt shortening with an elastic unloading (\( A\to B_{1} \)) followed by a massive conformational change in isometric conditions (\( B_{1}\to B_{2} \)). Parameters are, \( \lambda_{1} = 0.4 \), \( \lambda_0=0.7 \), \( l = -0.3 \), \( \lambda_{b} = 1 \), and \( N=3 \).}
	\label{fig:SS_zero_temperature_hard_device}
\end{figure*}
The energies of the equilibrium configurations can be now computed explicitly. For  a given \( (p,q,r) \) we obtain
\begin{equation}\label{eq:SS_v_hat}
	\hat{v}(p,q,r;z) =
	\frac{	\lambda_{b}\left[p\lambda_1(z+1)^{2}+q(z-l)^{2}+r\lambda_0z^{2}\right]
	+ p\lambda_1\left[r\lambda_0+q(1+2l)\right]-ql^{2}(q+\lambda_{b})
	}{
	2\left(\lambda_{b}+\Lambda(p,q,r)\right)
	}
	+ \frac{ql^{2}}{2(1-\lambda_0)}+(q+r)v_0
	\vphantom{\frac{\lambda_{b}^{2}}{()^{2}}}.
\end{equation}	
The  corresponding tension-elongation curves can be written as
\begin{equation*}
	\hat{t}\of{p,q,r;z} = \frac{\partial}{\partial z}\hat{v}(p,q,r;z)=\frac{\lambda_{b}\Lambda(p,q,r)}{\lambda_{b}+\Lambda(p,q,r)}\left(z+\frac{p\lambda_1-ql}{\Lambda(p,q,r)}\right).
\end{equation*}
Each triple  \( (p,q,r) \) defines an equilibrium branch which extends between the two limits  $[z_{\inf},z_{\sup}]$  induced by the inequalities \( \hat{x}_{1}<l \) and \( \hat{x}_{0}>l  \). We obtain
\begin{equation}\label{eq:SS_zsup_zinf}
	\begin{split}
	z_{\inf}(p,q,r) &= \frac{l\left[\lambda_{b}+\Lambda(p,q,r)\right]+(1-\lambda_0)(p\lambda_1-ql)}{\lambda_{b}(1-\lambda_0)},\\
	z_{\sup}(p,q,r) &= \frac{
	(l+\lambda_1)\left[
	\lambda_{b}+\Lambda(p,q,r)
	\right]+(1-\lambda_1)(p\lambda_1-ql)}{\lambda_{b}(1-\lambda_1)}.
	\end{split}
\end{equation}
The analysis presented in \ref{app:hard_device} shows that all equilibria with \( q=0 \) are stable while all the ones with  \( q\neq 0 \) are unstable. Therefore, as in the HS model, the metastable configurations in the snap-spring model can be parameterized by a single parameter \( p \). The obtained results are illustrated in  Fig.~\ref{fig:SS_zero_temperature_hard_device}.

We now show that the global minimum of the energy is again achieved exclusively on homogeneous configurations   $(1,0,0)$ and $\left(0,0,1\right)$. Assuming that \( q = 0, r=1-p \) and computing the second derivative of $v$ in (\ref{eq:SS_v_hat}) while interpreting $p$ as a continuous variable leads to
\begin{equation}
	\label{eq:SS_d2v_d2p}
 \frac{\partial ^{2}}{\partial p^{2}}\hat{v}\of{p;z} =-\frac{1}{2} \frac{\left[\lambda_{b}\lambda_1\left(z+1\right)+\lambda_0\left(\lambda_1-z\lambda_{b}\right)\right]^{2}}
 {\left[p\lambda_1+\left(1-p\right)\lambda_0+\lambda_{b}\right]^{3}}\leq0.
  \end{equation}
This inequality shows that the energy is concave, which means that the ground states are necessarily synchronized and separated by mixed configurations with higher energy levels forming a \emph{macroscopic} energy barrier. Observe that for \( \lambda_{b}\to\infty \) we have \( \frac{\partial ^{2}}{\partial p^{2}}\hat{v}\of{p;z}\to 0 \) and thus the macroscopic barrier disappears and we recover the degeneracy of the HS model.
The switch between the two homogeneous states takes place at the elongation \( z=z_{*} \) which solves \( \hat{v}(0,0,1;z_*) = \hat{v}(1,0,0;z_*) \).
We obtain
\begin{equation*}
	z_{*}=
	\begin{cases}
		\left[\left(\lambda_{0}-\lambda_1\right)\lambda_{b}\right]^{-1}
		\left[
			\lambda_{1}\left(\lambda_0+\lambda_{b}\right)
			-\sqrt{
			\mu
				\left(\lambda_0+\lambda_{b}\right)
				\left(\lambda_1+\lambda_{b}\right)
			}
		\right]
		&
		\text{if \( \lambda_1\neq\lambda_0 \)}\\
		\displaystyle{\frac{\lambda_1+\lambda_{b}}{\lambda_1\lambda_{b}}v_0 - \frac{1}{2}}
		&
		\text{if \( \lambda_1=\lambda_0 \)}
	\end{cases}
\end{equation*}
where \( \mu = \lambda_0\lambda_1+2(\lambda_1-\lambda_0)v_0 \geq 0 \). 
Notice that we recover the position of the transition point of the HS model, \( z_{*} = v_0-1/2 \), when considering the symmetric case with \( \lambda_{1}=1 \) and \( \lambda_{b}\to\infty \).

While the global minimum path in the snap-spring model  has the same structure as in the HS model, we see that at the transition point \( z=z_{*} \)  the energies of the mixture states are now strictly higher than the energies of the coexisting pure states, see Fig.~\ref{fig:SS_zero_temperature_hard_device}(a,c).  

The ensuing force-elongation relations, presented  in Fig.~\ref{fig:SS_zero_temperature_hard_device}(b,d), show  
  that the singular metamaterial behavior exhibited by the HS model in a hard device is not regularized in the snap-spring model where the stiffness corresponding to the globally stable response is still equal to  minus infinity at the transition point. 

\subsection{Mechanical equilibrium in a soft device} 
\label{sub:SS_soft_device}

To find equilibrium states in the  snap-spring  model loaded in a soft device we need to solve the system
	\begin{empheq}[left=\empheqlbrace]{align*}
		\left.\frac{\partial w}{\partial x_{i}}\right|_{t,z,y,\left\{x_{j\neq i}\right\}} &= 0\ \text{for all } 1\leq i \leq N \\
		\left.\frac{\partial w}{\partial y}\right|_{t,z,\left\{x_{i}\right\}} &= 0 \\
		\left.\frac{\partial w}{\partial z}\right|_{t,y,\left\{x_{i}\right\}} &= 0
	\end{empheq}
As in the hard device case, each crosslinker can be in three states and the equilibrium branches can be  parameterized by the triplet \( (p,q,r) \). After elimination of the internal degrees of freedom \( x_{i} \) and \( y \), the corresponding partially equilibrated energy can be written in the form
\begin{equation*}
	\tilde{w}(p,q,r,z;t) = \hat{v}(p,q,r,z) - tz
\end{equation*}
where the function \( \hat{v} \) is  given by Eq.~\ref{eq:SS_v_hat}.
After elimination of \( z \), the solution of the full mechanical equilibrium is obtained,
	\begin{align}
		\hat{y}(p,q,r;t) &=\frac{1}{\Lambda(p,q,r)}t - \frac{p\lambda_1-ql}{\Lambda(p,q,r)}, 
		\\
	 \hat{x}_0(p,q,r;t) &= \frac{(1-\lambda_0)}{\Lambda(p,q,r)}(t - p\lambda_1+ql),\label{eq:SS_hat_x0_of_p_sd}
	\\
	\hat{x}_1(p,q,r;t) &= \frac{(1-\lambda_1)}{\Lambda(p,q,r)}( t- p\lambda_1 + ql)-\lambda_1. \label{eq:SS_hat_x1_of_p_sd}
	\end{align}
	
The energy of a configuration $(p,q,r)$ can be again computed explicitly,
\begin{equation*}
	\hat{w}(p,q,r;t) = -\frac{1}{2}
	\left(\frac{1}{\lambda_{b}}+\frac{1}{\Lambda(p,q,r)}\right)t^{2}
	+ \frac{p\lambda_1-ql}{\Lambda(p,q,r)}t
	+ \frac{p\lambda_1r\lambda_0 - q^2l^2+2p\lambda_1ql}{2\Lambda(p,q,r)}
	+(q+r)v_0+\frac{ql^2}{1-\lambda_0},
\end{equation*}
and the corresponding tension-elongation relations read
\begin{equation*}
	\hat{z}\of{p,q,r;t} = -\frac{\partial}{\partial t}\hat{w}(p,q,r;t)
	= \left(\frac{1}{\lambda_{b}}+\frac{1}{\Lambda(p,q,r)}\right)t - \frac{p\lambda_1-ql}{\Lambda(p,q,r)}.
\end{equation*}
Similarly, we can obtain the lower and upper limits for  a branch labeled by $(p,q,r)$,
	\begin{equation}
		\label{eq:SS_tsup_tinf}
		\begin{split}
	 t_{\sup}(p,q,r) &= \frac{l+\lambda_1}{1-\lambda_1}\Lambda(p,q,r)+p\lambda_1-ql,\\
	t_{\inf}(p,q,r) &= \frac{l}{1-\lambda_0}\Lambda(p,q,r)+p\lambda_1-ql.
	\end{split}
	\end{equation}

The stability analysis presented in \ref{app:soft_device} shows again that configurations with \( q\neq 0 \) are unstable and therefore all metastable states can be parameterized by a single parameter \( p \).  Since
\begin{equation*}
	\frac{\partial^{2}}{\partial p^{2}}
\hat{w}\of{p;t}=
-\left[\frac{\left(\lambda_1-\lambda_0\right)^{2}\left(t-p\lambda_1\right)^{2}}{\left[p\lambda_1+\left(1-p\right)
\lambda_0\right]^{3}}
+\frac{\lambda_1^{2}}{p\lambda_1+\left(1-p\right)\lambda_0}
\vphantom{\frac{2\left(\lambda_1-\lambda_0\right)^{2}\left(t-p\lambda_1\right)^{2}}{\left(p\lambda_1+\left(1-p\right)\lambda_0\right)^{3}}}
\right]\leq0,
\end{equation*}
the global minimum is again attained either at \( p=1 \) or   \( p=0 \) and these pure states are robust and separated by a macroscopic energy barrier. Notice that in contrast to the hard device case, see Eq.~\ref{eq:SS_d2v_d2p}, the presence of this barrier does not depend on the backbone stiffness. The switch between them, signaling a collective folding-unfolding,  takes place   at \( t=t_* \) where \( \hat{w}(1,0,0,t_*) = \hat{w}(0,0,1,t_*) \). We obtain 
\begin{equation*}
	t_{*}=
		\left(\lambda_0-\lambda_1\right)^{-1}
		\left[
			\lambda_0\lambda_1-\sqrt{
				\mu\lambda_0\lambda_1
			}
		\right]
\end{equation*}
which simplifies into \( t_*=v_0 \) if \( \lambda_1=\lambda_0 \).
Note that in this later case, the value of \( t_{*} \) is the same as in the HS model.

	\begin{figure*}[htbp]
		\centering
		\includegraphics{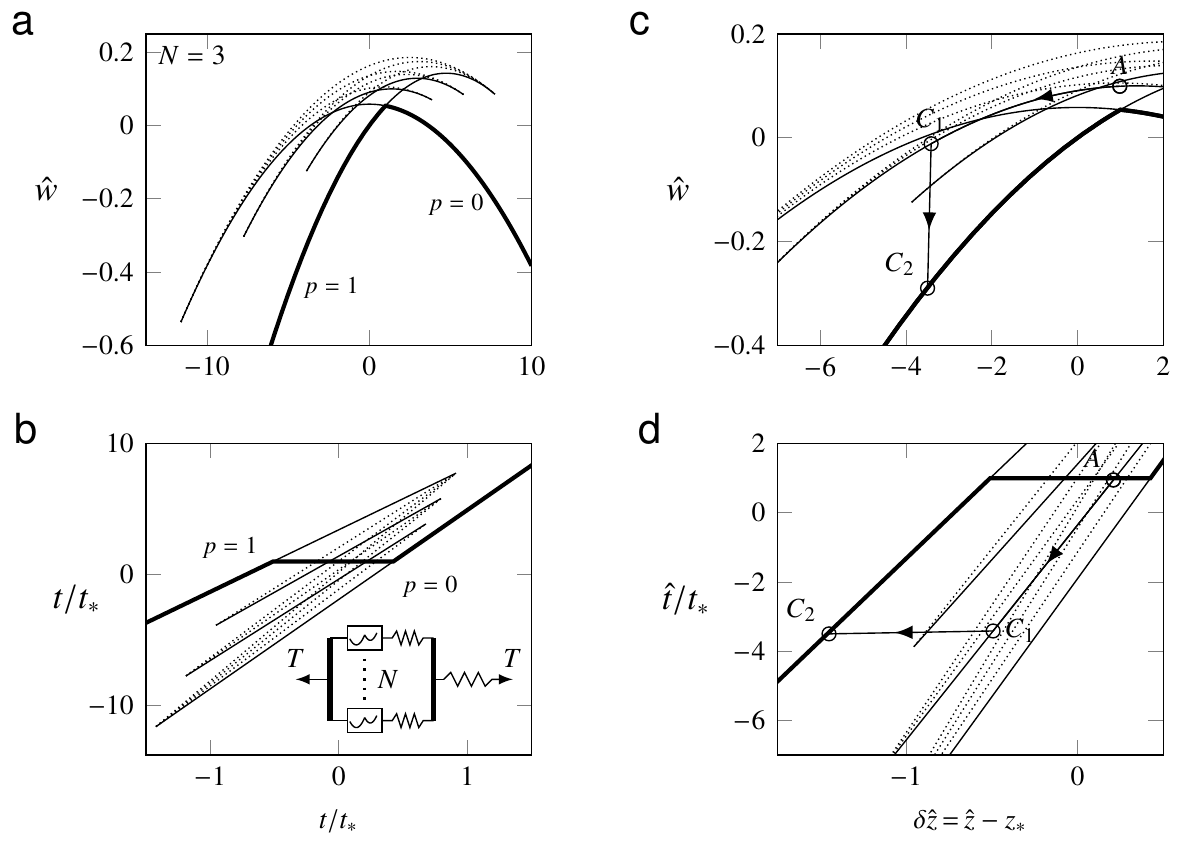}
		\caption{Mechanical response of the snap-spring model in a soft device. (a) Energy levels of all the \( (p,q,r) \) configurations for the case \( N=3 \) at different applied tensions. (b) corresponding tension-elongation relations. Solid lines, metastable states with \( p=0,1/3,2/3,1 \) and \( r=1-p \), \( q=0 \); dotted lines, unstable states with \( q\neq 0 \); thick lines, global minimum corresponding to \( p=0,\,r=1 \) for  \( t>t_* \) and to \( p=1,\,r=0 \), for \( t<t_* \). (c,d) zoom in of (a,b) illustrating the response of the system to abrupt shortening with an elastic unloading (\( A\to B_{1} \)) followed by a massive conformational change in isometric conditions (\( C_{1}\to C_{2} \)). Parameters are as in Fig.~\ref{fig:SS_zero_temperature_hard_device}.}
		\label{fig:SS_zero_temperature_soft_device}
	\end{figure*}

The equilibrium behavior of the snap-spring model in a soft device is illustrated  in Fig.~\ref{fig:SS_zero_temperature_soft_device}. As in  HS model,  the globally stable response contains an extended  plateau at \( t=t_* \) where the systems collectively switches between fully folded and fully unfolded configurations.  The constitutive behavior   remains quantitatively the same in the continuum limit $N \rightarrow \infty $. 


\subsection{Energy landscape } 
\label{sub:SS_energy_barrier}

We now study the size of the energy barrier separating the homogenous configurations characterized by \( p=0 \) and \( p=1 \). In the HS model, we have demonstrated that such macroscopic barrier exists only in the soft device case. In the snap-spring model the mechanical feedback introduced by the backbone elasticity introduces a variable degree of cooperativity between the crosslinkers which results in an increased energy of the mixed states already in a hard device as shown in Fig.~\ref{fig:SS_zero_temperature_hard_device}.

To find the energy minimizing ``reaction path'' connecting the homogeneous states through the set of inhomogeneous metastable states with $0<p<1$. The energy of these intermediate state can be made explicit by putting $q=0, r=1-p$  in  \eqref{eq:SS_v_hat}. We obtain,
	\begin{equation}
		\label{eq:SS_vhat_stable_state}
		\hat{v}(p;z) =
		\frac{1}{2\left(\lambda_b+\hat{\Lambda}(p)\right)}\left[
			\lambda_{b}\left(p\lambda_1(z+1)^{2}+(1-p)\lambda_0z^{2}\right)
		+
		p\lambda_1(1-p)\lambda_0\right]
		+(1-p)v_0,
	\end{equation}
	with 
	\begin{equation}
		\label{eq:SS_Lambda_hat}
		\hat{\Lambda}(p) = p\lambda_1+(1-p)\lambda_0.
	\end{equation}
We have shown that for $\lambda_{b} < \infty $ the function $\hat{v}(p;z)$ is concave in \( p \), see Eq.~\ref{eq:SS_d2v_d2p}, which signals the presence of a macroscopic energy barrier.
However, since  the variable $p$ is discrete, this information is incomplete and we need to also account for the \emph{microscopic} barriers separating configurations with different values of $p$.
Such barriers,  associated with the conformational changes in individual crosslinkers, were essentially infinite in the HS model.

To reconstruct the fine structure of the energy barriers, we consider a configuration with \( N_1 \) crosslinkers in the folded state, \( N_0 \) crosslinkers in the unfolded state and \( N_*=N-N_1-N_0 \) crosslinkers switching collectively from the unfolded to the folded state.
The initial stable configuration is fully characterized by the parameter \( p=N_{1}/N \) and we denote by \( \alpha = N_*/N \) the fraction of switching crosslinkers which satisfies \( 0\leq \alpha\leq 1-p \).

To find the barrier which the system has to overcome, we need to choose a microscopic ``reaction path'' separating the initial configuration characterized by  $p$ and  the final  configuration  characterized by  \( p+\alpha \).
Assuming for simplicity that the switching crosslinkers are characterized by the same strain variable \( x \), it is natural to choose \( x \) as a ``reaction coordinate''.

Due to permutational invariance, the choice of the \( N_* \) switching crosslinkers is arbitrary, and for commodity, we select \( x_{1},\dots,x_{N_*} \).
The energy landscape (per crosslinker) along the chosen reaction path can be recovered if we  minimize out the rest of the internal variables \(x_{N-N_*},\dotsc,x_{N},y \). We obtain
\begin{multline}
	\label{eq:SS_barrier_xn_hd}
	\bar{v}(p,\alpha,x;z)=
	\frac{1}{2}\left[
		\frac{
		\lambda_{b}\left(p\lambda_1(z+1)^{2}+(1-p-\alpha)\lambda_0z^{2} \right)
		+
		p\lambda_1(1-p-\alpha)\lambda_0
		}{
		\lambda_b+\hat{\Lambda}(p)+(1-\lambda_0)\alpha
		}
		\right]
		+(1-p-\alpha)v_0\\
		+
		\alpha\left[
			u_{\mbox{\tiny{SS}}}(x) + \frac{1}{2}
			\frac{
			\lambda_{b}(z-x)^2 + p\lambda_1(x+1)^2 + (1-p-\alpha)\lambda_0x^2
			}{
			\lambda_b+\hat{\Lambda}(p)+(1-\lambda_0)\alpha
			}
		\right],
\end{multline}
where the first term does not depend on \( x \); it  converges to \( \hat{v}(p;z) \) given by Eq.~\ref{eq:SS_vhat_stable_state} in the thermodynamic limit when \( N\to\infty \). The second term vanishes in the thermodynamic limit because it is proportional to \( \alpha \) which goes to zero when \( N\to\infty \).

For convenience, we map the reaction coordinate to the interval  \( [p,p+\alpha] \)  by replacing $x$  with a stretched variable \( \xi \) defined by
\begin{equation}\label{eq:SS_xi}
	\xi = p+\alpha\frac{x-\hat{x}_{0}(p;z)}{\hat{x}_{1}(p+\alpha;z)-\hat{x}_{0}(p;z)},
\end{equation}
where $\hat{x}_{0}$ and $\hat{x}_{1}$ are the locations of the bottoms of the energy wells defined by Eqs.\ref{eq:SS_hat_x1_of_p_hd} and \ref{eq:SS_hat_x0_of_p_hd}.
	One can see that  the variable \( \xi \) can be viewed as a \emph{global} reaction coordinate that encompasses the \emph{local}  	reaction coordinates of individual transitions in a step by step manner  and thereby defining the position of the system on the 	whole reaction path (see also \cite{Truskinovsky:2003iz}).

The values \( \xi=p \) and \( \xi=p+\alpha \) are associated with the metastable states \( \hat{v}(p,q=0,r=1-p;z) \) and \( \hat{v}(p+\alpha,q=0,r=1-p-\alpha;z) \), respectively, see Eq.~\ref{eq:hs_hd_energy_local_minima}. 
At \( \xi=p+\alpha(l-\hat{x}_0)/(\hat{x}_{1}-\hat{x}_0) \), we have    \( x=l \) and the energy has a local maximum, namely \( \hat{v}(p,q=\alpha,1-p-\alpha) \). These states belong to the unstable equilibrium branches  characterized by \( q>0 \)  and shown  in Fig.~\ref{fig:SS_zero_temperature_hard_device} by dotted lines.

	The local ``microscopic'' energy barriers surrounding a given metastable state characterized by  a particular \( p \) may be either in the direction of additional  folding (\( B_{\rightarrow} \)) or additional unfolding (\( B_{\leftarrow} \)) of a fraction \( \alpha \) of crosslinkers. The height of these barriers can be expressed analytically (from Eqs.~\ref{eq:SS_vhat_stable_state} and \ref{eq:SS_barrier_xn_hd} ) as follows
	\begin{equation}\label{eq:SS_HD_barrier_finite_N}
	\begin{split}
		B_{\rightarrow}(p,\alpha;z) &= N\left[\bar{v}(p,\alpha,l;z) - \hat{v}(p;z)\right],\\
		B_{\leftarrow}(p,\alpha;z) &= N\left[\bar{v}(p-\alpha,\alpha,l;z) - \hat{v}(p;z)\right].
	\end{split}
	\end{equation}
	In the thermodynamic limit, we have \( \alpha=N_{*}/N\to0 \) and the overall microscopic energy barriers take the form
	\begin{equation}\label{eq:SS_HD_barrier_asymptotic_developement}
	\begin{split}
		B_{\rightarrow}(p;z) &\sim a(p;z)-\frac{1}{N} b(p;z),\\
		B_{\leftarrow}(p;z) &\sim a'(p;z)-\frac{1}{N} b'(p;z).
	\end{split}
	\end{equation}
	The explicit expressions  for the coefficients \( a \), \( b \), \( a' \) and \( b' \) are given in \ref{sec:energy_barriers_in_a_hard_device}. We observe that in the thermodynamic limit, the height of the energy barriers between different metastable states has a finite limit while the height of a generic barrier per crosslinker (\( B_{\rightarrow}/N \) or \( B_{\leftarrow}/N \)) vanishes. Similar results have been previously obtained for chains of bistable elements connected in series \citep{Puglisi_2000,Benichou:2011fsa,Manca:2013ig,Tshiprut:2009kf}.

\begin{figure*}[htbp]
	\centering
		\includegraphics{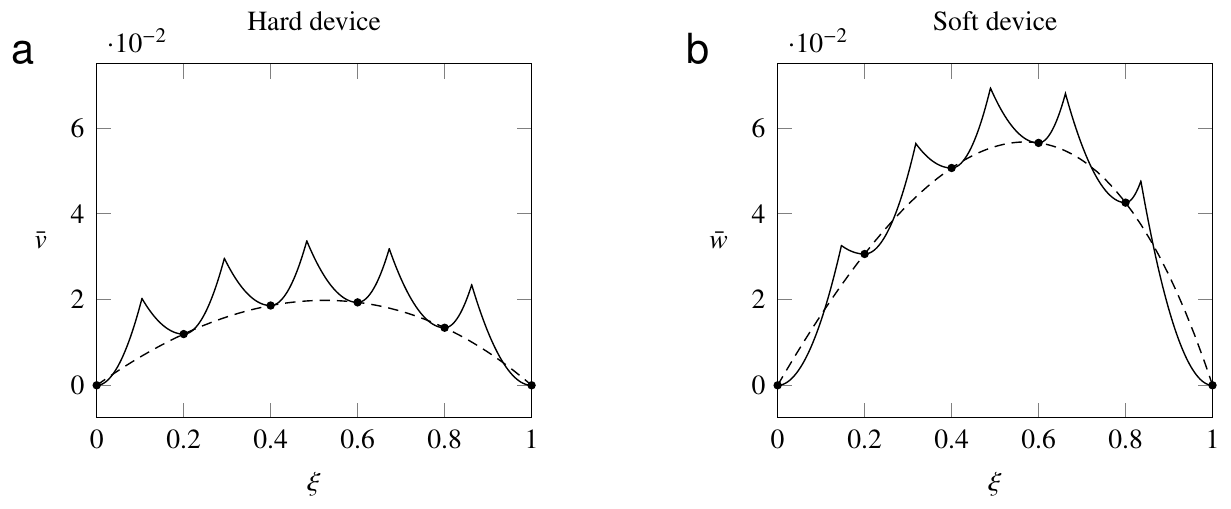}
	\caption{
		Energy landscape  at the global minimum transition for the snap-spring model with \( N=5 \).
		(a) hard device at \( z=z_{*} \); (b) soft device at \( t=t_{*} \).
		Solid lines, successive barriers obtained from Eqs.~\ref{eq:SS_barrier_xn_hd} (a) and \ref{eq:SS_wbar} (b); Dashed lines, continuum limit, \( N\to\infty \).
		Energy minima are arbitrarily set to 0 for comparison.
 		Here, \( \lambda_{b}=1 \) and other parameters are as in Fig.~\ref{fig:SS_zero_temperature_hard_device}. 
		In particular, we have \( \lambda_{0}\neq\lambda_1 \) which explains why the curve is not symmetric.
	}
	\label{fig:SS_zero_temperature_transition_single}
\end{figure*}

To facilitate comparison with the HS model we first limit our attention to the barriers associated with the transition between the two globally stable coherent states taking place  at \( z=z_{*} \) and consider only the energy minimizing reaction path which is caracterized by successive single crosslinker conformational change (\( \alpha=1/N \)).  The resulting energy landscape \( \bar{v}(\xi) \) at \( z=z_{*} \) is shown in Fig.~\ref{fig:SS_zero_temperature_transition_single}(a) where we compare two cases, \( N=5 \) (solid line and metastable states marked with dots)  and  \( N\to\infty \) (dashed line).
At finite \( N \) we see the \emph{macroscopic barrier}, not captured by the HS model and a superimposed set of \emph{microscopic barriers}   representing  ``lattice pinning''. These microscopic barriers  are due to the discreteness of the problem and they disappear in the continuum/thermodynamic limit \( N=\infty \) when \(\xi \to p\).\\

In the case of a soft device,
	the energy of metastable states is obtained by inserting \( q=0 \) in Eq.~\ref{eq:hs_sd_energy_local_minima}.
	 We obtain
	\begin{equation}
		\label{eq:SS_what}
		\hat{w}(p;t) = -\frac{1}{2}
		\left(\frac{1}{\lambda_{b}}+\frac{1}{\hat{\Lambda}(p)}\right)t^{2}
		+ \frac{p\lambda_1}{\hat{\Lambda}(p)}t
		+ \frac{p\lambda_1(1-p)\lambda_0}{2\hat{\Lambda}(p)}
		+(1-p)v_0.
	\end{equation}
	The partially equilibrated energy at fixed \( t \), \( p \), \( \alpha \) and \(x\) is then given by
	\begin{multline}
		\label{eq:SS_wbar}
		\bar{w}(p,\alpha,x;t) = -\frac{1}{2}
		\left(\frac{1}{\lambda_{b}}+\frac{1}{\hat{\Lambda}(p) + (1-\lambda_0)\alpha}\right)t^{2}
		+ \frac{
		p\lambda_1(1-p-\alpha)\lambda_0 
		+ 2p\lambda_1t
		}{2(\hat{\Lambda}(p) + (1-\lambda_0)\alpha)}
		+(1-p-\alpha)v_0\\
		+\alpha\left[u_{\mbox{\tiny{SS}}}(x)
		+\frac{1}{2}\frac{
		p\lambda_1(x+1)^{2} + (1-p-\alpha)\lambda_0x^{2}-2tx
		}
		{
		\hat{\Lambda}(p) + (1-\lambda_0)\alpha
		}
		\right].
	\end{multline}
	We again map the reaction coordinate \( x \) in the interval \( [p,p+\alpha] \) using Eq.~\ref{eq:SS_xi} where \( \hat{x}_{0}(p;z) \) and \( \hat{x}_{1}(p;z) \) have to be replaced by \( \hat{x}_{0}(p;t) \) and \( \hat{x}_{1}(p;t) \) given by Eq.~\ref{eq:SS_hat_x0_of_p_sd} and Eq.~\ref{eq:SS_hat_x1_of_p_sd}, respectively.	
	 The microscopic energy barriers for folding (\( B_{\rightarrow} \)) or unfolding (\( B_{\leftarrow} \)) of \( N_{*} \) crosslinkers   are obtained analytically from Eqs.~\ref{eq:SS_what} and \ref{eq:SS_wbar},
	\begin{equation}\label{eq:SS_SD_barrier_finite_N}
	\begin{split}
		B_{\rightarrow}(p,\alpha;t) &= N\left[\bar{w}(p,\alpha,x;t) - \hat{w}(p;t)\right],\\
		B_{\leftarrow}(p,\alpha;t) &= N\left[\bar{w}(p-\alpha,\alpha,x;t) - \hat{w}(p;t)\right].
	\end{split}
	\end{equation}
	As in the hard device case, we obtain in the thermodynamic limit,
	\begin{equation}\label{eq:SS_SD_barrier_asymptotic_developement}
		\begin{split}
		B_{\rightarrow}(p;t) &\sim c(p;t)-\frac{1}{N} d(p;t),\\
		B_{\leftarrow}(p;t) &\sim c'(p;t)-\frac{1}{N} d'(p;t).
	\end{split}
	\end{equation}
	The explicit formulas for the coefficients \( c \), \( c' \), \( d \) and \( d' \) are given in \ref{sec:energy_barrier_in_a_soft_device}; as one can expect, these coefficients do not depend on the backbone stiffness \( \lambda_b \).
	Notice that the result is fully analogous to what we obtained in the case of a hard device, see Eq.~\ref{eq:SS_HD_barrier_asymptotic_developement}: in the thermodynamic limit, the individual overall energy barriers remain finite while the barriers per cross-bridge vanish.
The fine structure of the ensuing energy  landscape (per cross-bridge) is illustrated in Fig.~\ref{fig:SS_zero_temperature_transition_single}(b)  for  \( t=t_{*} \) and successive transitions with \( \alpha=1/N \). At finite \( N \), we see again the  two-scale structure, however, the macroscopic barrier is markedly higher in a soft than in a hard device.  
\begin{figure*}[tbp]
	\centering
		\includegraphics{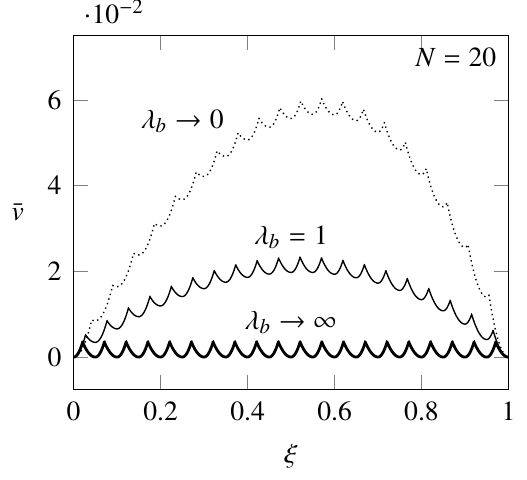}
	\caption{Energy landscape at the global minimum transition for the snap-spring model in a hard device at different values of the coupling parameter \( \lambda_{b} \) with \( N=20 \). Bold line, \( \lambda_{b}\to\infty \); solid line, \( \lambda_{b} = 1 \); dotted line, soft device limit as the hard device case where \( \lambda_{b}\to 0 \), \( z_{*}\to\infty \). Energy minima are arbitrarily set to 0 for comparison. Other parameters are as in Fig.~\ref{fig:SS_zero_temperature_hard_device}.}
	\label{fig:SS_zero_temperature_transition}
\end{figure*}

In  Fig.~\ref{fig:SS_zero_temperature_transition} we illustrate the dependence of the energy landscape in a hard device on the parameter \( \lambda_{b} \) characterizing the backbone elasticity in a system with \( N=20 \). To simplify the comparison we adjusted the parameter \( z=z_{*} \) at each value of \( \lambda_{b} \) so that the coherent states with \( p=0 \) and \( p=1 \) have the same energy (arbitrarily set to zero for the sake of comparison). As we know, the snap-spring model converges to the HS model as \( \lambda_{b} \rightarrow \infty \) and we see in Fig.~\ref{fig:SS_zero_temperature_transition} that the ``macroscopic barrier'' disappears in this limit. The microscopic barriers remain and that is what distinguishes the model proposed in \citep{Marcucci_2010} from the HS model. Of course, in the continuum limit \( N \rightarrow \infty \)  we loose the microscopic barriers and recover the interpolated version of the HS model, see Fig.~\ref{fig:HS_zero_temperature_transition}(a).

In the same figure we illustrate the limit \( \lambda_{b} \rightarrow 0 \) and  \(  z_{*} \rightarrow \infty \).  In this case we recover in the hard device setting the predictions of the  soft device model with \( t \rightarrow t_{*} \). Once again, in the continuum limit \( N \rightarrow \infty \) the microscopic barriers disappear and we recover the basic picture predicted by the HS model.

Similar analysis of the barrier structure  can be performed for a generic loading. 
Consider first the hard device case. In the continuum limit \( N\to\infty \), the variable \( p \) becomes  continuous. We denote by \( p_{*}(z) \) the position of energy barrier (saddle point) which is obtained by solving \( \partial\hat{v}/\partial p=0 \) for a given loading \( z \). We obtain
\begin{equation*}	
	p_{*}(z) =
	\begin{cases}
		\displaystyle{
		 \frac{
		\sqrt{\mu}(\lambda_0+\lambda_{b}) + \lambda_0\lambda_{b}z
		-\lambda_1\lambda_{b}(z+1)-\lambda_0\lambda_1
		}{
		\sqrt{\mu}(\lambda_0-\lambda_1)
		}}
		& \text{if \( \lambda_1\neq\lambda_0, \)}\\[10pt]
		\displaystyle{
		\frac{
		\lambda_1[\lambda_1 + \lambda_{b}(1+2z)] - 2v_0(\lambda_1+\lambda_{b})
		}{2\lambda_1^{2}}
		}
		& \text{if \( \lambda_1=\lambda_0 \)}.
	\end{cases}
\end{equation*}
Since \( 0\leq p_*\leq 1 \),  the macroscopic energy barrier exists only if \( z_{-}<z<z_{+} \), where
\begin{align*}
	z_{-} &=
	\begin{cases}
		\left[
			\lambda_{b}
			\left(
				\lambda_0-\lambda_1
			\right)
		\right]^{-1}
		\left[
			\left(\lambda_1 - \sqrt{\mu}\right)
			\left(\lambda_0+\lambda_{b}\right)
		\right] 
		& 
		\text{if \( \lambda_{1}\neq\lambda_0 \)}\\[5pt]
		\displaystyle{
		\frac{
			\left(
				2v_0-\lambda_1
			\right)
			\left(
				\lambda_1 + \lambda_{b}
			\right)
		}{
			2\lambda_1\lambda_{b}
		}
		}
		&
		\text{if \( \lambda_{1}=\lambda_0 \)}
	\end{cases}\\
	z_{+} &=
	\begin{cases}
		\left[
			\lambda_{b}
			\left(
				\lambda_0-\lambda_1
			\right)
		\right]^{-1}
		\left[
			\lambda_1
			\left(\lambda_0+\lambda_{b}\right)
			-\sqrt{\mu}
			\left(
				\lambda_0+\lambda_{b}
			\right)
		\right]
		&\text{if \( \lambda_{1}\neq\lambda_0 \)}\\[5pt]
		\displaystyle{
		\frac{
		2v_0\left(\lambda_1+\lambda_{b}\right)
		+ \lambda_1\left(\lambda_1-\lambda_{b}\right)
		}{2\lambda_1\lambda_{b}}
		}
		&
		\text{if \( \lambda_{1}=\lambda_0 \)}
	\end{cases} 
\end{align*}

Notice that for an arbitrary \( z \), the reaction path does not have to pass through all intermediate configurations, see Eq.~\ref{eq:SS_zsup_zinf}.  By inverting Eq.~\ref{eq:SS_zsup_zinf}, we obtain that for a given elongation, the available metastable states are characterized by \( p_{\inf}(z)<p<p_{\sup}(z) \) with,
\begin{equation}\label{eq:SS_psup_pinf_hd}
	\begin{split}
		p_{\sup}(z) &=  \frac{\lambda_{b} z(1 - \lambda_{0}) - l(\lambda_{0}+\lambda_{b})}{
l(\lambda_{1}-\lambda_{0}) + (1 - \lambda_{0})\lambda_1},\\
		p_{\inf}(z) &=  \frac{\lambda_{b} z(1 - \lambda_{1}) - (l+\lambda_1)(\lambda_{0}+\lambda_{b})}{
(l+\lambda_1)(\lambda_{1}-\lambda_{0}) + (1 - \lambda_{1})\lambda_1}.\\
	\end{split}
\end{equation}
Therefore, during the transition the system explores all metastable states satisfying Eq.~\ref{eq:SS_psup_pinf_hd} before cooperatively switching directly to one of the homogeneous state with \( p=1 \) or \( p=0 \).
\begin{figure}[!ht]
	\centering
	\includegraphics{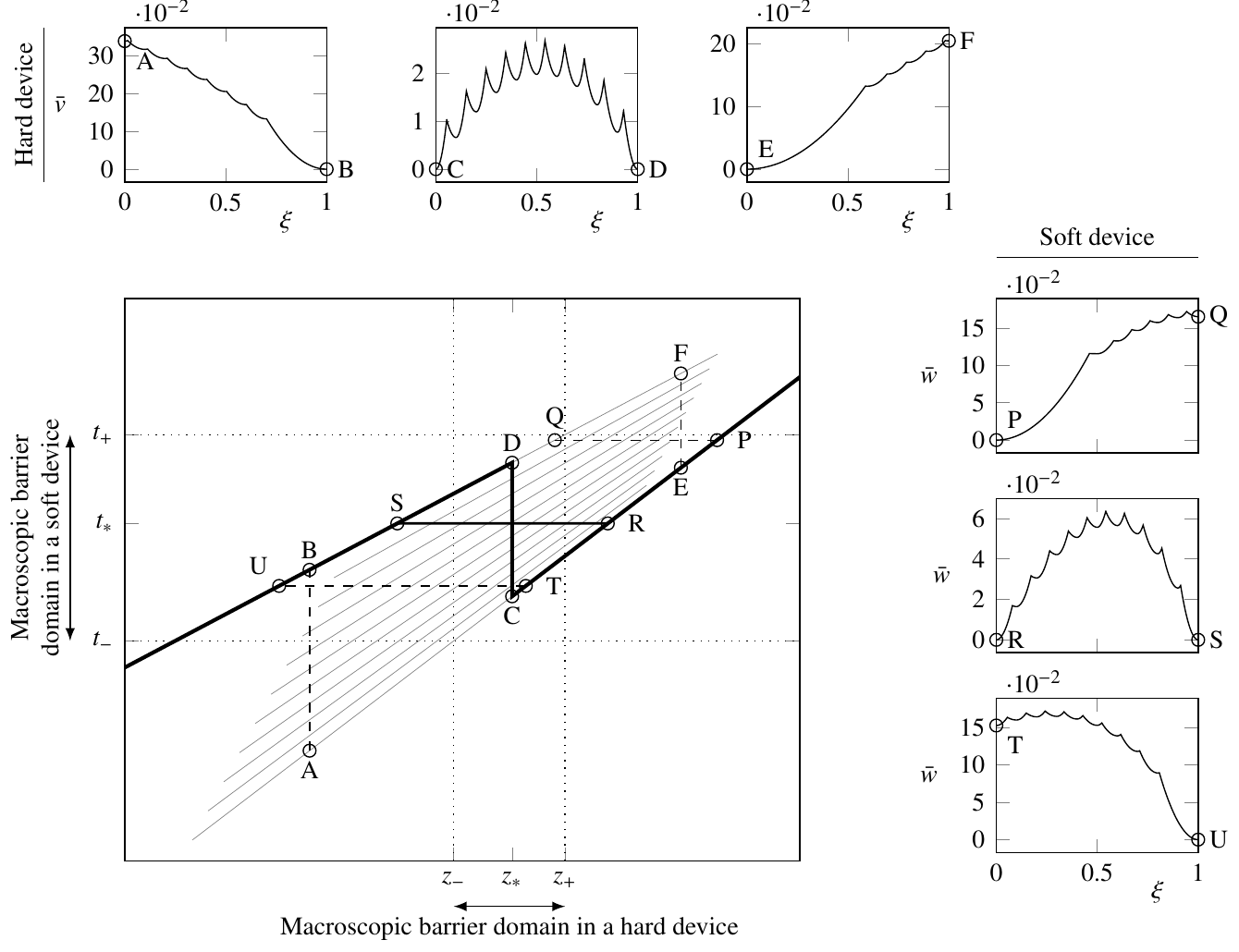}
	\caption {\label{fig:SS_conclusion_figure}Overall behavior of the snap-spring model. The main frame shows the tension-elongation relations corresponding to \( p=0,0.1,\dots,1\), for a system with \( N=10 \) (gray lines) and the tension-elongation relations in the global minimum (thick lines). The dotted lines show the limit of the domain where a macroscopic energy barrier is present in a hard device (vertical lines) and in a soft device (horizontal lines). The satellite frames (above, hard device; right, soft device) show the energy barriers corresponding to various transitions shown by dashed lines in the main frame.
		In particular, the transition C\( \to \)D (hard device) and the transition R\( \to \)S (soft device) correspond to \( z=z_{*} \) and \( t=t_{*} \), respectively. The other hard device transitions, A\( \to \)B and E\( \to \)F are outside the bistable interval \( [z_{-},z_{+}] \) and thus contain a collective transition. The other soft device transitions, P\( \to \)Q and T\( \to \)U exhibit the same kind of synchronization along the segment of the 'reaction path' which is located inside the bistable domain.
	Parameters other than \( N \) are as in  Fig.~\ref{fig:SS_zero_temperature_hard_device}.}
\end{figure}

These results are summarized in Fig.~\ref{fig:SS_conclusion_figure} where the main frame shows the tension-elongation relations corresponding to the different metastable states for a system with \( N=10 \). The satellite frames show the transition path related landscape corresponding to three different values of \( z \). For instance, in the case where \( z<z_{*} \) (see A\( \to  \)B) all metastable states with \( p\leq7/10 \) are available and then the system coherently switches from the state with \( p=7/10 \) to the state with \( p=1 \). On the main frame, we indicate by vertical dotted lines the values of \( z_{-} \) and \( z_{+} \) which delimit the domain where the macroscopic barrier exists.

A similar analysis can be conducted in the soft device case. First, considering the continuum limit \( N\to\infty \), we solve \( \partial\bar{w}/\partial p=0 \) to locate the position of the energy barrier  \( p_{*} \). We obtain,
\begin{equation*}
	p_{*}(t) = 
	\begin{cases}
		\left[\left(\lambda_0-\lambda_1\right)\sqrt{\mu}\right]^{-1}
		\left[
			\left(\lambda_1-\lambda_0\right)t
			+\lambda_0\left(\lambda_1+\sqrt{\mu}\right)
		\right]
		&
		\text{if \( \lambda_1\neq\lambda_0 \)}\\[5pt]
		\displaystyle{
		\frac{t-v_0}{\lambda_1} + \frac{1}{2}
		}
		&
		\text{if \( \lambda_1=\lambda_0 \)}.
	\end{cases}
\end{equation*}
Then the condition \( 0\leq p_{*}(t)\leq 1 \) gives the interval \( [t_{-},t_{+}] \) where the macroscopic barrier is present.  Here
\begin{align*}
	t_{-} &= 
	\begin{cases}
		\left[\lambda_0-\lambda_1\right]^{-1}
		\left[
			\lambda_0
			\left(
				\lambda_1 + \sqrt{\mu}
			\right)
		\right]
		&
		\text{if \( \lambda_{0}\neq\lambda_1 \)}\\
		v_0 - \lambda_{1}/2
		&
		\text{if \( \lambda_{0}=\lambda_1 \)}
	\end{cases}\\
	t_{+} &= 
	\begin{cases}
		\left[\lambda_0-\lambda_1\right]^{-1}
		\left[
			\lambda_1
			\left(
				\lambda_0 + \sqrt{\mu}
			\right)
		\right]
		&
		\text{if \( \lambda_{0}\neq\lambda_1 \)}\\
		v_0 + \lambda_{1}/2
		&
		\text{if \( \lambda_{0}=\lambda_1 \)}
	\end{cases}
\end{align*}
Notice that we recover the same boundaries as in the HS model if we consider the limit of infinite stiffness \( \lambda_1\to 1 \) and set \( \lambda_1=\lambda_0 \).
Finally, for a given \( t \), the available metastable states satisfy \( p_{\inf}(t)\leq p\leq p_{\sup}(t) \). The boundaries  can be obtained by inverting Eq.~\ref{eq:SS_tsup_tinf}
\begin{align*}
		p_{\sup}(t) &=  
		\frac{t (\lambda_0-1) + l \lambda_0}
		{l \lambda_0 + (\lambda_0-l-1) \lambda_1},
		\\
		p_{\inf}(t) &= 
		 \frac{t (\lambda_1-1) + \lambda_0 (l + \lambda_1)}{
		l \lambda_0 + (\lambda_0-l - 1) \lambda_1}.
	\end{align*}
Again this shows that (i) the interval where the macroscopic barrier is present is finite and that (ii) the transition between the globally stable homogenous states may involve coherent  switches depending on the value of the applied tension. We illustrate these results in Fig.~\ref{fig:SS_conclusion_figure}, where the satellite plots on the right show the energy barrier as function of \( \xi \)  and where the horizontal dotted lines delimit the macroscopic barrier domain.

We finally remark that while  the individual local minima of the energy landscape at finite $N$  can be interpreted as distinct   \emph{chemical states}, the exponentially growth of the number of such states in the thermodynamic limit makes this  interpretation unpractical. Instead, the description in terms of  macro-wells and the associated \emph{mechanisms} can lead to a meaningful quasi-chemical representation of the loading-induced unfolding process.  In the case of multiscale unfolding processes, the macro-wells corresponding to  \emph{synchronized configurations} are usually hierarchically structured which makes the mechanical description of such systems an interesting challenge.

	\section{Application to skeletal muscles} 
	\label{sec:application_to_the_case_of_skeletal_muscles}
	\begin{table}
		\centering
		\begin{tabular}{llp{1cm}ll}
			 \toprule
			 \multicolumn{2}{c}{Dimensional}&&
			 \multicolumn{2}{c}{Non-dimensional}\\
			 \midrule
			 \( a\)				&	\(10\,\mbox{nm} \)				&&					&					\\
			 \( l \)			&	\( -0.8\, \mbox{nm} \)			&&	\( l \)			&	\( -0.08 \)		\\
			 \( k \)			&	\(2.7\, \mbox{pN.nm}^{-1}\)		&&	\( N \)			&	\( 100 \)		\\	
			 \( k_{b} \)		&	\( 135\, \mbox{pN.nm}^{-1} \)	&&	\( \lambda_b \)	&	\( 0.5 \)		\\
			 \( k_{1} \)		&	\(0.3\, \mbox{pN.nm}^{-1}\)		&&	\( \lambda_1 \)	&	\( 0.23 \)		\\
			 \( k_{0} \)		&	\(0.8\, \mbox{pN.nm}^{-1}\)		&&	\( \lambda_0 \)	&	\( 0.44 \)		\\
			 \( k_{b}\theta \)	&	\( 3.8\times 10^{-21} \,\mbox{J} \)	&&					&					\\
			 \bottomrule
		\end{tabular} 
		\caption{Realistic parameters for the snap-spring model applied to skeletal muscles.}
		\label{tab:SS_parameters}
	\end{table}

	In this section, we briefly discuss how our results can be  applied  to skeletal muscles. Our Table~\ref{tab:SS_parameters} summarizes the realistic values of parameters calibrated in \citep{Caruel:2013jw}. We chose \( N=100 \)  by taking into consideration that only one third of cross-bridges are attached at any given moment of time \citep{Piazzesi_2007}.

\begin{figure}[!ht]
		\centering
		\includegraphics{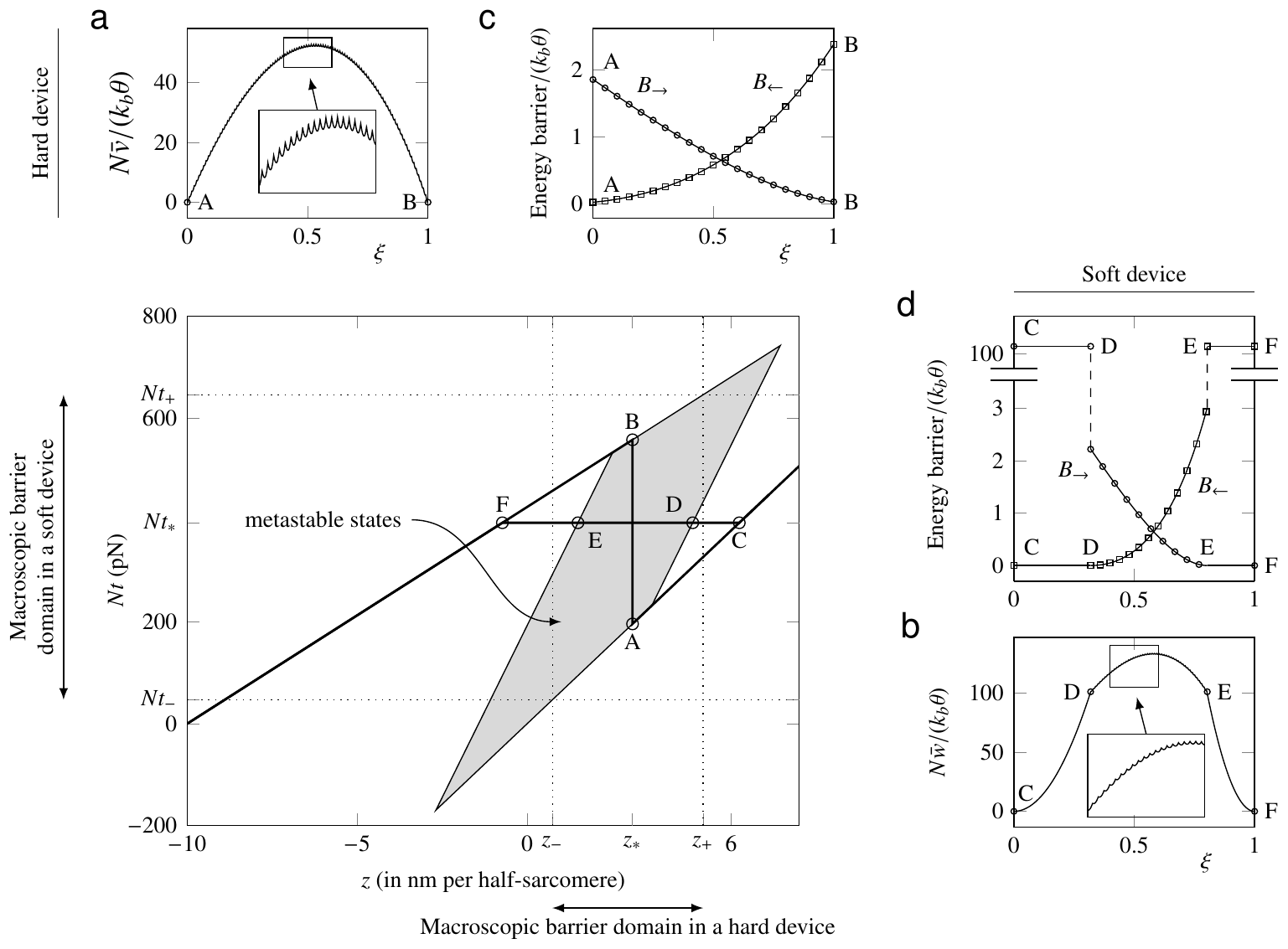}
		\caption{\label{fig:real_case}Result of the soft spin model with parameters adjusted to fit experimental data, see \cite{Caruel:2013jw}.
		Main frame, tension-elontation relation for a single half-sarcomere. The gray area shows the existence domain of the metastable states. Solid lines, global minimum tension-elongation relation in hard and soft devices. (a) and (b), Energy landscape corresponding to single transition between the homogeneous states in a hard device (A\( \to \)B), see (a) and in a soft device (C\( \to \)F), see (b). (c) and (d) size of the energy barriers corresponding to the individual folding (\( B_{\rightarrow} \)) and (\( B_{\leftarrow} \)) transitions obtained with both finite \( N \) (open symbols) and in the thermodynamic limit (solid lines). To improve readability of the finite \( N \) plots, we only show one out of five micro-transition points.}
	\end{figure}

	In Fig.~\ref{fig:real_case} we show the metastability domain (gray area) and the tension-elongation relations describing the global minimum of the energy. In the satellites frames  we illustrate two  sections of the energy landscape along the paths connecting the two homogeneous states: in a hard device at \( z=z_{*} \), see Fig.~\ref{fig:real_case} (a), and in a soft device at \( t=t_{*} \), see Fig.~\ref{fig:real_case} (b). These plots are obtained by computing
	the energy \( N\bar{v} \), see Eq.~\ref{eq:SS_barrier_xn_hd} and the energy \( N\bar{w} \), see Eq.~\ref{eq:SS_wbar} for hard and soft device, respectively. In addition we show the size of the individual energy barriers for single folding (\( B_{\rightarrow} \)) and unfolding (\( B_{\leftarrow} \)) events, see Fig.~\ref{fig:real_case} (c and d).  The results  for   finite \( N \) (open symbols) were obtained by using Eq.~\ref{eq:SS_HD_barrier_finite_N} for the hard device case and Eq.~\ref{eq:SS_SD_barrier_finite_N} for the soft device case.  In the thermodynamic limit (solid lines) we used Eq.~\ref{eq:SS_HD_barrier_asymptotic_developement} in the hard device case and Eq.~\ref{eq:SS_SD_barrier_asymptotic_developement} in the soft device case.  Observe that the zone of bistable behavior in a soft device, is much broader that in a hard device and spans almost the entire metastability domain. 
	This correlates with the fact that the macroscopic energy barrier in a soft device, see Fig.~\ref{fig:real_case}(b) is about three times higher than in a hard device, see Fig.~\ref{fig:real_case}(a). 

Since experiments on muscle fibers are performed at finite temperature, it is of interest to compare the height of the microscopic and macroscopic energy barriers with the typical energy of thermal fluctuations \( k_{b}\theta \) where \( k_{b} \) is the Boltzmann constant and \( \theta \) is the absolute temperature (see similar analysis of microscopic barriers for titin in \cite{Benichou:2011fsa}).
	
In a hard device, the full range of metastable states between the two homogeneous states (labeled by A and B) is available, see Fig.~\ref{fig:real_case} (a). In Fig.~\ref{fig:real_case} (c) we show the size of the microscopic barriers corresponding to single folding (resp. unfolding) events, see \( B_{\rightarrow} \) (resp. \( B_{\leftarrow} \)). These energy barriers are of the order of \( k_{b}\theta \) which suggests that thermal noise alone can unfold single bistable elements. However, at small values of \( \xi \) the height of the energy barriers corresponding to folding is systematically higher than for unfolding   and \emph{vice-versa} for large  values of  \( \xi \). Therefore, although  individual folding/unfolding transitions can occur due to thermal fluctuations, the presence of a macroscopic barrier, which is at least 50 times higher, ensures that the system is maintained in  globally synchronized states, here corresponding to points A or B.

In a soft device, see Fig.~\ref{fig:real_case} (b and d), the situation is  slightly different.  One can see that at \( t=t_{*} \), the reaction path passes through a reduced fraction of metastable states with high energy  in the interval \( 0.3\leq\xi\leq0.8 \). Therefore, to switch from one  homogeneous configurations to another (say, from C to F), the system must first reach the metastable states D or E through a massive collective transition (C\( \to \)D or F\( \to \)E) requiring an energy of about \( 100 \,k_{b}\theta\), see Fig.~\ref{fig:real_case} (b and d). These massive transitions are very unlikely, moreover  the reverse transitions have almost zero energy barrier. Once the system has reached the metastable states, D or E, we are back to the pattern encountered in the case of  a hard device   where energy barriers for individual transitions were of the order of \( k_{b}\theta \). The fact that the macroscopic barrier in the case of soft device is much higher than in the case of hard device  may be the reason behind the anomalously slow kinetics of relaxation observed in isotonic experiments \citep{Piazzesi_2002, Reconditi_2004,Decostre_2005}.

To summarize, individual contractile units in skeletal muscles can be viewed as an assembly of nanometer sized bi-stable mechanisms. In the absence of long-range coupling, these mechanisms  would transform individually producing strongly inhomogeneous  temporal and spatial microstructures. The presence of long-range interactions is a way for the system to strongly bias homogeneous states and  in this way passively synchronize individual bistable mechanisms. The transition rate associated with a macroscopic transition is extremely sensitive to the number of switching elements as the (macroscopic) energy barrier is proportional to \( N \). This suggests that the system can fine-tune its kinetics by recruiting a particular number of crosslinkers. Our computation of the individual barrier heights,  using the asymptotic formulas given by Eqs.~\ref{eq:SS_HD_barrier_asymptotic_developement} and \ref{eq:SS_SD_barrier_asymptotic_developement}, shows that the energy per crossbridge in the asymptotic regime of infinitely large \( N \) is already well approximated at the realistic values \( N\sim100 \). This means that this number of cross-bridges, characterizing a single half-sarcomere, is sufficient for the system to achieve the maximum synchronization ability.  Even though our conclusions  are reached based on the  analysis of the internal  rather than the free energy, they are fully confirmed  by the detailed study of the finite temperature effects on muscle dynamics based on the direct modeling of  the Langevin dynamics \citep{Caruel:2013jw}.
 

\section{Conclusions} 
\label{sec:conclusion}

A  parallel bundle of bistable snap-springs is  a  simple mechanical system which, however, teaches important lessons.  The presence in this system of mean-field type interactions, induced by the coupling of individual bistable units through a common backbone, produces a peculiar mechanical behavior. For instance, the transition between the two states  (folded and unfolded)   takes the form of a collective switching event rather than a sequence of transitions in individual elements. The system exhibits negative stiffness  and, even in the continuum limit, the  mechanical behavior  is different in soft and hard loading devices.  Such systems,  where each element is linked with almost equal strength with all other elements and  the whole is not a sum of the parts are ubiquitous in biology with skeletal muscles providing just one of the many examples.  While in such systems the individual units may be submitted to random thermal fluctuations, the domineering   long-range interactions  provide a highly efficient way to maintain the individual units  passively synchronized. More specifically,  long-range interactions  impose a strong bias between forward and backward reaction rates for microscopic transitions favoring globally ordered states.
 
The   prototypical  model studied in this paper has important applications outside the skeletal muscle context. We have already mentioned  the phenomena of synchronized unzipping of adhesive clusters \citep{Erdmann_2007,Chen:2011go,Yao:2006de,Gao:2011ij,Erdmann:2013wr}   and   the  cooperative  flip-flopping  of macro-molecular hairpins \citep{Liphardt:2001fp,Prados:2012ks,Bosaeus:2012kp,Woodside:2008uz}. A more complex but related example  is provided by the  ``fracture'' avalanches during unfolding of macro-molecules  \citep{Srivastava:2013tm}. The broad applicability of the proposed  mechanical perspective is also corroborated by the fact that proteins and nucleic acids  behave differently in isometric and isotonic conditions and that these mechanical systems can exhibit negative stiffness \citep{Gerland:2003ie,Bornschlogl:2006we,Thomas:2012ur}. In the same vein, the importance of the topology of interconnections among the bonds and the link between the cooperativity of unfolding and the dominance of parallel bonding have been long stressed in the studies of protein folding \citep{Dietz:2008gj}.  By emphasizing the crucial role of the force transmitting backbones our study provides a  simple paradigmatic description of this class of phenomena. On a more practical side, the  model suggests an explicit path  towards designing  bio-mimetic materials and molecular nano-machines whose functioning depends essentially on long-range feedback between multi-stable units \citep{Yurke:2000tw}.

\bigskip

\section{Acknowledgements} 

The authors are grateful to P. Recho, R. Sheshka and V. Lombardi for helpful discussions and to the anonymous reviewer for constructive suggestions.

\bigskip
\appendix

\section{Stability in a hard device}
\label{app:hard_device}

The  analysis of stability is similar in the HS and the snap-spring models. A technical complication is that in both cases
the  bistable potentials $u_{\mbox{\tiny HS}}$ and  $u_{\mbox{\tiny SS}}$ are singular.  Thus, in the HS model the energy wells are infinitely narrow  and the energy barrier is formally infinite. In the snap-spring model, the energy wells have a finite curvature however the spinodal region is reduced to a single point. To study stability, we first remove these singularities by considering smoother potentials and then perform the appropriate limiting transition to singular potentials.

We start with the more general snap-spring model and regularize the bistable potential $u_{\mbox{\tiny SS}}$ by introducing an extended spinodal interval \(\left[l-\epsilon;l+\epsilon\right]\) where the new potential $\tilde{u}_{\mbox{\tiny SS}}$ is concave.  Assume that outside this interval the   potential $\tilde{u}_{\mbox{\tiny SS}}$  coincides with $u_{\mbox{\tiny SS}}$  and is  therefore convex. In a hard device the new snap-spring energy  can be written in the form
\begin{equation}\label{SS_energy_app}
 v(\boldsymbol{x},y;z)=\frac{1}{N} \sum_{i=1}^{N} \left[\tilde{u}_{\mbox{\tiny SS}}(x_{i})+ \frac{1}{2}(y-x_{i})^{2}\right] +\frac{\lambda_{b}}{2}(z-y)^{2}.
\end{equation}

We now analyze stability of the system described by energy \eqref{SS_energy_app}. At a given $z$, the equilibrium equations can be written as
\begin{equation}\label{SS_energy_app1}
	\frac{\partial v}{\partial x_{i}} = \tilde{u}'_{\mbox{\tiny SS}}(x_{i}) + (x-\hat{y})=0,\quad i=1,\dots,N,
\end{equation}
where \( \hat{y}(z,p,q,r) \) is  the equilibrium value of \( y \) for a given configuration \( (p,q,r) \)  given by Eq.~\ref{eq:SS_hat_y_of_p_hd}.
Assume that within the spinodal region there is an interval where \( \tilde{u}''_{\mbox{\tiny SS}}(x)<-1 \). Then each of the  equations \eqref{SS_energy_app1} has up to 3 solutions. We denote these solution by $\hat{x}_1(z,p,q,r),\hat{x}_0(z,p,q,r)$ and $\hat{x}_{\star}(z,p,q,r)$.
The first two solutions \( \hat{x}_{1} \) and \( \hat{x_{0}} \) correspond to the two convex wells of the potential  $\tilde{u}_{\mbox{\tiny SS}}(x)$  so that \( \hat{x}_{1}<l-\epsilon \) and \( \hat{x}_{0}>l+\epsilon \).  The third solution $\hat{x}_{\star}$ describes the crosslinker in the spinodal region and therefore $l-\epsilon\leq\hat{x}_{*}\leq l+\epsilon$ for all $\epsilon>0$.  We can now compute
\begin{align*}
 \left.\frac{\partial^{2}v(\boldsymbol{x},y;z)}{\partial x_{i}^{2}}
\right|_{p,q,r,x_i=\hat{x}_{1},y=\hat{y},x_{j\neq i}=\hat{x}_{j}}
&=\k_{1}+1\equiv h_{1}>0,
\\
\begin{split}
 \left.\frac{\partial^{2}v(\boldsymbol{x},y;z)}{\partial x_{i}^{2}}
\right|_{p,q,r,x_i=\hat{x}_{*},y=\bar{y},x_{j\neq i}=\hat{x}_{j}}
&=\tilde{u}''_{\mbox{\tiny SS}}(\hat{x}_{*})+1\\
&\equiv h_{*}(z,p,q,r)<0,
\end{split}\\
 \left.\frac{\partial^{2}v(\boldsymbol{x},y;z)}{\partial x_{i}^{2}}
\right|_{p,q,r,x_i=\hat{x}_{0},y=\hat{y},x_{j\neq i}=\hat{x}_{j}}
&=\k_{0}+1\equiv h_{0}>0.
\end{align*} %
Here \( h_{*} \) is negative because the corresponding crosslinker is the spinodal state. The other second derivatives of the energy can be computed explicitly%
\begin{align}
	\nonumber
&\left.\frac{\partial^{2}v(\boldsymbol{x},y;z)}{\partial x_{i}\partial x_{j}}
\right|_{p,q,r,x_i=\hat{x}_{i},x_{j}
=\hat{x}_{j},x_{k\neq i,j}=\hat{x}_{k},y=\hat{y}}
=0\,\, \mathrm{for}\ i\neq j,\\
\label{eq:hessian_other}
&\left.\frac{\partial^{2}v(\boldsymbol{x},y;z)}{\partial x_{i}\partial y}
\right|_{p,q,r,x_i=\hat{x}_{i},x_{j\neq i}
=\hat{x}_{j},y=\hat{y}}
=-1,\,\, \ i=1,\dots,N,
\\
\nonumber
&\left.\frac{\partial^{2}v(\boldsymbol{x},y;z)}{\partial y^{2}}
\right|_{p,q,r,x_i=\hat{x}_{i},y=\hat{y}}
 = 1+\lambda_{b}.
\end{align}%

To write the expression for the Hessian matrix $\textbf{H}(z,p,q,r)$, it is  convenient to introduce the following auxiliary quantities  
\begin{equation*}
 H_i(p,q,r;z) =  \left.\frac{\partial^{2}v(\boldsymbol{x},y,z)}{\partial x_{i}^{2}}
\right|_{p,q,r,x_i=\hat{x}_{i},y=\hat{y},x_{j\neq i}
=\hat{x}_{j}}.
\end{equation*}
Each of the variables  $H_i$ can  take  three  values: $h_1$, $h_0$ and $h^{\star}$.  Now we can write
\begin{equation}\label{eq:SS_hessian}
\textbf{ H}(p,q,r;z)=
\left(
   \begin{matrix}
      H_{1}&0&\cdots&0&-1\\
      0&\ddots&\ddots&\vdots&\vdots\\
      \vdots&\ddots&\ddots&0&\vdots\\
     0&\cdots&0&H_{N}&-1\\
      -1&\hdotsfor{2}&-1&1+\lambda_{b}	
   \end{matrix}
\right).
\end{equation}%
To obtain a similar Hessian matrix for the HS model, we need to perform the limit  \( \k_{1,0}(z)\to\infty \) which means \( h_{1}\to\infty \), \( h_{0}\to\infty \) and \( h_{*}\to -\infty \), and to drop in \eqref{eq:SS_hessian} the last line and the last column.

From the form of the matrix $\textbf{H}$, one can see that as soon as one of the terms $H_i$ is equal to $h^{\star}(z)$, which means that \( q>0 \), at least one of the principal minors of $\textbf{H}$ becomes negative. This means that the absence of crosslinkers in the spinodal region is mandatory for stability  at $\epsilon>0$. 

We can now consider the limit  $\epsilon\to 0$. The value of the equilibrium strain in the spinodal region \( \hat{x}_{*} \) remains between $l-\epsilon$ and $l+\epsilon$ and thus converges to $l$, when $\epsilon\to 0$. Therefore, if $q\neq0$, the configuration $\left(p,q,r\right)$ is necessary unstable and we know that such configurations are necessarily singular.
Hence, in our snap-spring model, among the $\left(N+1\right)\left(N+2\right)/2$ equilibrium branches, $N\left(N+1\right)/2$ singular branches are unstable which leaves $N+1$ nonsingular branches describing local minima of the energy. Since in the HS model the spinodal states are absent, all configurations are automatically metastable.

\section{Stability in a soft device}
\label{app:soft_device}
In the soft device case, the tension \( t \) is fixed while \( z \) becomes an additional degree of freedom. Then the  energy of the nonsingular snap-spring system reads
\begin{equation}\label{SS_energy_app_sd}
 w(\boldsymbol{x},y,z;t)=\frac{1}{N} \sum_{i=1}^{N} \left[\tilde{u}_{\mbox{\tiny SS}}(x_{i})+ \frac{1}{2}(y-x_{i})^{2}+\frac{\lambda_{b}}{2}(z-y)^{2} - tz\right] .
\end{equation}

The analysis of the equilibrium states remains the same and we can similarly define the diagonal terms of the Hessian matrix for the energy $w(\boldsymbol{x},y,z;t)$,
\[
 H_{i}(p,q,r;t) =   \left.\frac{\partial^{2}w(\boldsymbol{x},y,z,t)}{\partial x_{i}^{2}}
\right|_{p,q,r,x_i=\hat{x}_{i},y=\hat{y},z=\hat{z},x_{j\neq i}
=\hat{x}_{j}}.
\]
Here each term can take the following three values $h_1=\k_{1}+1>0$,  $h_0(p,q,r,t)=\k_{0}+1>0$ or $h^{\star}(p,q,r,t)<0$. The other entries are the same as in the  hard device case, see Eq.~\ref{eq:hessian_other} except that now we have one  additional row and one additional column,
\begin{align*}
	\begin{split}
&\left.\frac{\partial^{2}w(\boldsymbol{x},y,z;t)}{\partial x_{i}\partial z}
\right|_{p,q,r,x_i=\hat{x}_{i},x_{j\neq i}
=\hat{x}_{j},y=\hat{y},z=\hat{p,q,r;t}}
=0,\\&\quad\quad\ \text{for } i=1,\dots,N
\end{split}\\
&\left.\frac{\partial^{2}w(\boldsymbol{x},y,z;t)}{\partial y\partial z}
\right|_{p,q,r,x_i=\hat{x}_{i},x_{j\neq i}
=\hat{x}_{j},y=\hat{y},z=\hat{z}}
=-\lambda_{b},\\
&\left.\frac{\partial^{2}w(\boldsymbol{x},y,z;t)}{\partial z^{2}}
\right|_{p,q,r,x_i=\hat{x}_{i},x_{j\neq i}
=\hat{x}_{j},y=\hat{y},z=\hat{z}}
=\lambda_{b}.
\end{align*} %
By bringing all these second derivatives together we can write  the Hessian matrix for the snap-spring model in a soft device 
\begin{equation*}
\textbf{ H}(p,q,r;t)=\
 \left (
   \begin{matrix}
      H_{1}&0&\cdots&0&-1&0\\
      0&\ddots&\ddots&\vdots&\vdots&\vdots\\
      \vdots&\ddots&\ddots&0&\vdots&\vdots\\
     0&\cdots&0&H_{N}&-1&0\\
      -1&\hdotsfor{2}&-1&\left(1+\lambda_{b}\right)&-\lambda_{b}	\\
	0&\hdotsfor{2}&0&-\lambda_{b}&\lambda_{b}
   \end{matrix}
   \right ).
\end{equation*}%
A straightforward  adaptation of the above analysis shows that, as in a hard device, the system in a soft device is unstable only when \( q\neq0 \), \emph{i.e.} when at least one cross-bridge is in the spinodal state.

Finally,  to obtain the Hessian matrix for the HS system we need to drop the last row and the last column and consider the limit  \( \k_{1,0}(z)\to\infty \) which means \( h_{1}\to\infty \), \( h_{0}\to\infty \) and \( h_{*}\to -\infty \). We also require that \( \lambda_{b} \to 0 \). Then, the Hessian reads
\begin{equation*}
 \textbf{H}_{HS}(p,q,r;t)=\
 \left (
   \begin{matrix}
      H_{1}&0&\cdots&0&-1\\
      0&\ddots&\ddots&\vdots&\vdots\\
      \vdots&\ddots&\ddots&0&\vdots\\
     0&\cdots&0&H_{N}&-1\\
      -1&\hdotsfor{2}&-1&1\\
   \end{matrix}
   \right ).
\end{equation*}%
The analysis here is similar to the case of a hard device and the conclusion is that again all equilibrium configurations are metastable.
\section{Energy barriers in a hard device} 
\label{sec:energy_barriers_in_a_hard_device}
The explicit expressions for the coefficients in the the asymptotic development \eqref{eq:SS_HD_barrier_asymptotic_developement} are
\begin{align*}
	a(p;z) &= N_*
		\left[u_{\mbox{\tiny{SS}}}(l) - v_{0}
			+ \frac{
			\left\{\left[\lambda_b+\hat{\Lambda}(p)\right]l-(\lambda_bz-p\lambda_1)\right\}^{2} - 			\lambda_0\left[\lambda_bz-p\lambda_1\right]^{2}
			}{2\left[\lambda_b+\hat{\Lambda}(p)\right]^{2}}
		\right]\\
	b(p;z) &= N_{*}^{2}
		\frac{
		\left\{\left[\lambda_b+\hat{\Lambda}(p)\right]l-(1-\lambda_0)(\lambda_bz-p\lambda_1)\right\}^{2}
		}{2\left[\lambda_b+\hat{\Lambda}(p)\right]^{3}}\\
	a'(p;z) &= N_*
		\left[u_{\mbox{\tiny{SS}}}(l)
			+ \frac{
			\left\{\left[\lambda_b+\hat{\Lambda}(p)\right](l+1)-\left[\lambda_b(z+1)+(1-p)\lambda_0\right]\right\}^{2} - 			\lambda_1\left[\lambda_b(z+1)+(1-p)\lambda_0\right]^{2}
			}{2\left[\lambda_b+\hat{\Lambda}(p)\right]^{2}}
		\right]\\
	b'(p;z)&= N_{*}^{2}
		\frac{
		\left\{\left[\lambda_b+\hat{\Lambda}(p)\right](l+1)-(1-\lambda_1)\left[\lambda_b(z+1)+(1-p)\lambda_0\right]\right\}^{2}
		}{2\left[\lambda_b+\hat{\Lambda}(p)\right]^{3}}
\end{align*}
with \( \hat{\Lambda}(p) \) given by Eq.~\ref{eq:SS_Lambda_hat}.

\section{Energy barrier in a soft device} 

\label{sec:energy_barrier_in_a_soft_device}
The explicit expressions for the coefficients in the the asymptotic development \eqref{eq:SS_SD_barrier_asymptotic_developement} are
\begin{align*}
	c(p;t) & = 	N_*
		\left[u_{\mbox{\tiny{SS}}}(l) - v_{0} +
			\frac{
			\left(\hat{\Lambda}(p)l - (t-p\lambda_1)\right)^{2} - \lambda_0(t-p\lambda_1)^{2}
			}{
			2\hat{\Lambda}(p)^{2}
			}
		\right]\\
	d(p;t) & = 	N_*^{2}
			\frac{
			\left(\hat{\Lambda}(p)l - (1-\lambda_0)(t-p\lambda_1)\right)^{2}
			}{
			2\hat{\Lambda}(p)^{3}
			}\\
	c'(p;t) & = 	N_*
		\left[u_{\mbox{\tiny{SS}}}(l) +
			\frac{
			\left(\hat{\Lambda}(p)(l+1) - (t+(1-p)\lambda_0)\right)^{2} 
			- \lambda_1\left[t+(1-p)\lambda_0\right]^{2}
			}{
			2\hat{\Lambda}(p)^{2}
			}
		\right]\\
	d'(p;t) & = N_*^{2}
			\frac{
			\left(\hat{\Lambda}(p)(l+1) - (1-\lambda_1)(t+(1-p)\lambda_0)\right)^{2}
			}{
			2\hat{\Lambda}(p)^{3}
			}\\
\end{align*}
with \( \hat{\Lambda}(p) \) given by Eq.~\ref{eq:SS_Lambda_hat}.

\section*{\refname}
\bibliography{biblio}
\bibliographystyle{model2-names}


\end{document}

%% file: Declarations.tex
\usepackage[utf8]{inputenc}

\usepackage{txfonts}
\usepackage{graphicx}
\usepackage{tikz}

\usepackage{pgfplots}
\pgfplotsset{compat=newest}
\usetikzlibrary[pgfplots.groupplots]
\usetikzlibrary{arrows}
\usetikzlibrary{decorations.markings}

\usetikzlibrary{backgrounds}
\usetikzlibrary{chains}
\usetikzlibrary{calc}
\usetikzlibrary{external}
\usepackage{bm}
\usepackage{hyperref}

\usepackage{empheq}
\usepackage{xspace}
\usepackage{color}
\usepackage{amssymb}
\usepackage{booktabs}
\usepackage{appendix}

\hypersetup{
bookmarksnumbered = true,
unicode=false,          
pdftoolbar=true,        
pdfmenubar=true,        
pdffitwindow=false,     
pdfstartview={Fit},    
pdftitle={Passive force recovery in skeletal muscles I. Mechanical model},    
pdfauthor={M. Caruel},     
pdfsubject={Subject},   
pdfcreator={Creator},   
pdfproducer={Producer}, 
pdfkeywords={keyword1} {key2} {key3}, 
pdfnewwindow=true,      
colorlinks=true,       
linkcolor=black,          
citecolor=black,        
filecolor=black,      
urlcolor=black,           
pdfdisplaydoctitle = true
}

\renewcommand{\bar}{\overline}

\def\k{k}

\newcommand{\of}[1]{\left(#1\right)}

\usepackage[normalem]{ulem}
\usepackage{color}
\definecolor{myorange}{rgb}{0.9568,0.4941,0.1961}
\definecolor{myred}{rgb}{0.9098,0.1294,0.2078}
\definecolor{myblue}{rgb}{0.0352,0.4981,0.6509}
\definecolor{mygreen}{rgb}{0.2235,0.6353,0.2588}

\usepackage{marvosym}

\newcommand{\inserted}[1]{{\color{myblue}{{#1}}}}

